\documentclass[
  aps,pra,reprint,superscriptaddress,amsmath,amssymb,longbibliography,floatfix,raggedbottom
]{revtex4-2}

\usepackage{graphicx}
\usepackage{bm}
\usepackage{mathbbol}
\usepackage[T1]{fontenc}
\usepackage{lmodern}
\usepackage[colorlinks,linkcolor=blue,citecolor=blue,urlcolor=blue]{hyperref}
\newcommand{\ASTRA}{\textsc{astra}}
\newcommand{\XCHEM}{\textsc{xchem}}

\makeatletter
\newenvironment{widefigure}{\par\onecolumngrid\noindent\begin{minipage}{\textwidth}\def\@captype{figure}\centering}{\end{minipage}\par\vspace{8pt}\twocolumngrid}
\newenvironment{widetable}{\par\onecolumngrid\noindent\begin{minipage}{\textwidth}\def\@captype{table}}{\end{minipage}\par\vspace{8pt}\twocolumngrid}
\makeatother

\begin{document}

\title{Variational surface matching for multichannel molecular photoionization}

\newcommand{\ucf}{Department of Physics \& CREOL, University of Central
Florida, Orlando, Florida 32816, USA}
\newcommand{\uam}{Departamento de Qu\'imica, Universidad Aut\'onoma de Madrid,
28049 Madrid, Spain}

\author{Carlos Marante}
\affiliation{\ucf}

\author{Vicent J. Borr\`as}
\affiliation{\uam}

\author{Jesús González Vázquez}
\affiliation{\uam}

\author{Fernando Mart\'in}
\affiliation{\uam}

\author{Luca Argenti}\email{luca.argenti@ucf.edu}
\affiliation{\ucf}

\date{\today}

\begin{abstract}
We present an improved method for computing multichannel scattering states in the \ASTRA{} (AttoSecond TRAnsitions) molecular ionization code. Our pole-free Galerkin method with variational surface matching (PFG-VSM) yields a boundary response that is stationary with respect to errors in the interior wave function. Compared with matching based on directly evaluated radial derivatives, this formulation yields substantially faster convergence and greater numerical stability as the number of open channels increases. Comparison with a numerical solution of the exactly separable $\mathrm H_2^+$ problem establishes the accuracy attainable in a hybrid Gaussian--B-spline basis for a polycentric potential, independently of many-electron close-coupling approximations. A direct comparison of the two matching prescriptions for CO demonstrates the stability of the variational formulation under radial-basis coarsening. For neon, agreement with the NewStock atomic ionization code tests the treatment of the continuum degeneracies that arise when rotational symmetry is represented in $D_{2h}$. The excellent agreement with the independent \XCHEM{} molecular ionization code for CO partial cross sections and molecular-frame photoelectron angular distributions, both away from and across resonances, is a strong confirmation of the correctness of both codes. A comparison for pyrazine extends this validation to a larger polyatomic target.
\end{abstract}

\maketitle

\section{Introduction\label{sec:Introduction}}

Single-photon ionization is one of the most direct probes of electronic structure and of correlated electron dynamics in atoms and molecules. Beyond the integral cross section, the angular distribution of the emitted photoelectron encodes the relative amplitudes and phases of the interfering partial-wave ionization channels, and it does so with increasing detail as the frame of reference is refined: from the asymmetry parameters of the laboratory-frame photoelectron angular distribution (LFPAD) of a randomly oriented sample to the molecular-frame photoelectron angular distributions (MFPADs) accessible through coincidence detection of photoelectrons and fragment ions~\cite{Golovin1997,Lafosse2000,Lafosse2002,Dowek2004,Reid2012,Dowek2022} or through laser-induced alignment and orientation~\cite{Larsen2000,Seideman2005,Koch2019,Holmegaard2010}. MFPADs resolve the orientation-dependent interference structures that rotational averaging obscures, and they have become a standard tool to image molecular potentials, shape resonances, and, more recently, attosecond photoemission time delays in the molecular frame~\cite{Fukuzawa2019,Rist2021,Joseph2023,Ueda2019}. Supporting this class of experiments requires sophisticated \emph{ab initio} methods capable of computing energy- and channel-resolved complex ionization amplitudes with controlled accuracy in correlated many-electron systems.

For polyatomic molecules, the computation of these amplitudes is particularly challenging: the polycentric many-electron problem requires a balanced, simultaneous description of correlated bound and continuum states, including the correlated states of the residual cation and a photoelectron that explores both the molecular region and the long-range Coulomb tail~\cite{Martin2026}. Several successful approaches exist, which differ in how they represent the electronic continuum: the Schwinger multichannel method~\cite{Lucchese1986}, single-center scattering approaches such as ePolyScat~\cite{Gianturco1994,Natalense1999}, multicenter B-spline approaches such as Tiresia~\cite{Toffoli2024}, the complex Kohn variational method~\cite{McCurdy1987,Rescigno1995}, molecular $R$-matrix theory in its UKRmol+ implementation~\cite{Tennyson2010,Masin2020,Burke2011}, and hybrid Gaussian--B-spline close-coupling methods such as \XCHEM{}~\cite{Marante2017,Borras2021,Borras2024}, which extended to molecules the multi-channel scattering approach in the NewStock atomic code~\cite{Luo2024,Wood2026}. Time-dependent ionization can also be treated with the hybrid antisymmetrized coupled-channels (haCC) method implemented in tRecX~\cite{Chundayil2024}. Each of these methods trades differently between the sophistication of the target description, the flexibility of the continuum representation, and the scaling with system size.

\ASTRA{} (AttoSecond TRAnsitions)~\cite{Randazzo2023} is a recent addition to this landscape, designed to push the close-coupling approach towards larger polyatomic systems. Its distinctive feature is the evaluation of all many-electron matrix elements between close-coupling channels (antisymmetrized products of correlated ionic states and one active electron) by means of transition density matrices (TDMs) between general configuration-interaction states. Since the TDMs are computed once, the cost of assembling the close-coupling Hamiltonian does not scale with the size of the configuration-interaction expansion of the ionic states, which can therefore be described with larger active spaces. The active electron is represented in a hybrid basis of polycentric Gaussian orbitals, which are flexible in the molecular region, and of radial B-splines~\cite{Bachau2001}, which provide an accurate and numerically stable radial description of the continuum up to large distances, with the mixed integrals evaluated with the GBTOlib library~\cite{Masin2020,GBTOlib-4}. In its first applications, \ASTRA{} demonstrated total photoionization cross sections obtained from the optical theorem, i.e., from the imaginary part of the expectation value of the resolvent, computed by adding a complex absorbing potential (CAP) to the Hamiltonian~\cite{Randazzo2023,Siegert1939,McCurdy1979}. The scalar optical-theorem response used in those applications does not resolve the complex amplitudes of the individual outgoing channels. Those amplitudes are needed for partial cross sections and angular distributions.

The main result of the present work is a pole-free Galerkin method with variational surface matching (PFG-VSM) for the multichannel scattering problem. Its boundary response is stationary with respect to errors in the interior wave function. Compared with matching based on directly evaluated radial derivatives, this formulation yields substantially faster convergence with respect to the one-electron numerical basis and greater numerical stability of the scattering observables as the number of open channels increases. The implementation within ASTRA of this method makes these advantages available for molecular photoionization calculations.
We assess the scattering method through a sequence of complementary comparisons. The exactly separable $\mathrm H_2^+$ problem establishes its accuracy in a polycentric potential without many-electron approximations, and a CO discretization study compares the stability of direct and variational matching. Neon then tests the molecular implementation in  $D_{2h}$ symmetry, which leads to degenerate box eigenstates, against the calculations with the NewStock atomic ionization code, which exploits the full spherical symmetry of the system. For CO, the independent \ASTRA{} and \XCHEM{} calculations agree remarkably well in partial cross sections and MFPADs, both away from and across resonances. Their comparison for pyrazine extends the test to a larger polyatomic molecule.

Section~\ref{sec:Methods} develops the scattering construction and the photoionization observables. Section~\ref{sec:Results} presents the comparisons in the order outlined above, and Sec.~\ref{sec:Conclusions} summarizes their implications. Appendix~\ref{app:OrientationalAverage} derives the orientationally averaged observables, and Appendix~\ref{app:H2plusBenchmark} describes the prolate-spheroidal reference solution of $\mathrm H_2^+$. Atomic units ($\hbar=m_e=|q_e|=1$) and Gaussian electromagnetic units are used throughout unless otherwise stated.

\section{Methods\label{sec:Methods}}

The calculations presented here use real basis functions and real field-free Hamiltonian and overlap matrices. The scattering construction, however, applies equally to Hermitian Hamiltonians with complex matrix elements, such as those arising in the presence of spin--orbit coupling, and we develop it in this general form. 

\subsection{The close-coupling space\label{subsec:CCSpace}}

\ASTRA{} represents the $N_e$-electron states of the molecule in a close-coupling (CC) space of antisymmetrized products of correlated ionic states and active-electron orbitals, supplemented by localized channels. For the spin-free calculations considered here, the \emph{partial-wave-channel} (PWC) function~\cite{Randazzo2023} couples the parent-ion state $|A\rangle$ and the photoelectron spin to a total spin $S\Sigma$, with electron angular character $(\ell,m)$,
\begin{equation}\label{eq:PWCDefinition}
\Upsilon_{\alpha}[u] \equiv
\hat{\mathcal{A}}\,
\Big[\Phi_{A}\otimes{}^2\chi\Big]_{S\Sigma}
X_{\ell m}(\hat{r})\,\frac{u(r)}{r},
\end{equation}
where $X_{\ell m}$ is a real spherical harmonic, $^2\chi_\sigma(\zeta) = \delta_{\sigma\zeta}$, with $\sigma,\zeta\in\{\pm1/2\}$, are doublet spin functions, and $u(r)$ is the reduced radial function. The channel index is $\alpha=(A,\ell,m)$ at fixed total spin and spatial symmetry $\Gamma$ in the symmetry group used in the calculation, which is normally chosen as the largest common subgroup of the molecular invariance group and $D_{2h}$. The ionic states $\Phi_A$ are obtained by configuration-interaction calculations at the complete-active-space (CAS) level~\cite{Randazzo2023}. We use $\hat{\mathcal A}=\sqrt{N_e}\,\hat A_{N_e}$, with $N_e$ the number of electrons and $\hat A_{N_e}$ the idempotent antisymmetrizer~\cite{Randazzo2023}. 

The active-electron basis combines polycentric Gaussian orbitals and radial B-splines in the molecular region, $r<R_{\mathrm{mol}}$, with the B-spline basis extended beyond this region to a much larger matching radius \(R_{\mathrm{box}}\)~\cite{Randazzo2023,Bachau2001}. This radius is chosen so that interchannel coupling can be neglected beyond it. To improve the description of short-range correlation, we extend the close-coupling expansion to excited ionic states beyond those needed to represent the relevant ionization channels. For these additional states, the active-electron basis comprises either Gaussian orbitals alone or the hybrid Gaussian–B-spline basis within the molecular region, without the external B-splines. Optical-theorem calculations use the same radial box, with complex absorbing potentials supported between $R_{\textsc{cap}}$ and $R_{\mathrm{box}}$, where $R_{\textsc{cap}}> R_{\mathrm{mol}}$ the molecular region and the matching surface.

Continuum couplings are assembled from the one- and two-body ionic TDMs and hybrid-basis integrals. The fully internal block is supplied by the quantum-chemistry calculation. This separates the cost of assembling continuum couplings from the size of the ionic CI expansions.
For each channel whose B-spline basis extends to \(R_{\mathrm{box}}\), the last radial B-spline \(B(r)\) is the only one nonzero at that boundary. We denote the corresponding PWC function by \(|B_\alpha\rangle\) and call it a surface function. Here $B(r)$ is $L^2$-normalized.

\subsection{A complete set of scattering states\label{subsec:ScatteringStates}}

At real total energy $E$ above the first ionization threshold, we seek a complete set of linearly independent eigenfunctions of the field-free Hamiltonian $\hat H$ with specified scattering boundary conditions. The incoming-wave states $|\Psi^-_{\alpha E}\rangle$ are labeled by the open channel $\alpha$, with ionic threshold $E_A$, and satisfy
\begin{equation}\label{eq:SchrodingerContinuum}
(E-\hat H)|\Psi^-_{\alpha E}\rangle = 0,\qquad \forall\,\alpha \,|\, E_A<E,
\end{equation}
The asymptotic behavior, in channel $\beta$ and for the radial coordinate of the outermost electron, is
\begin{equation}\label{eq:IncomingBC}
\Psi^-_{\alpha E}
\;\underset{r\to\infty}{\longrightarrow}\;
\sum_\beta^{\mathrm{open}}\Upsilon_\beta
\left[\frac{e^{i\theta_\beta(r)}\delta_{\beta\alpha}
-e^{-i\theta_\beta(r)}[\mathbf S^\dagger]_{\beta\alpha}}
{\sqrt{2\pi k_\beta}}\right],
\end{equation}
where $k_\beta=\sqrt{2(E-E_B)}$ for a channel $\beta$ with ionic threshold $E_B$ and electron orbital angular momentum $\ell_\beta$, $\mathbf S$ is the scattering matrix, and
\begin{equation}\label{eq:CoulombPhase}
\theta_\beta(r)=k_\beta r+\frac{Z}{k_\beta}\ln(2k_\beta r)
-\frac{\ell_\beta\pi}{2}+\sigma_{\ell_\beta}(k_\beta),
\end{equation}
with $\sigma_\ell(k)=\arg\Gamma(\ell+1-iZ/k)$ and residual charge $Z$~\cite{Newton}. These states are normalized as
\begin{equation}\label{eq:EnergyNormalization}
\langle\Psi^-_{\alpha E}|\Psi^-_{\beta
E'}\rangle=\delta_{\alpha\beta}\,\delta(E-E'),
\end{equation}
so that $|\langle\Psi^-|\Psi\rangle|^2$ is an energy probability density. To determine these states, we first construct a complete set of stationary solutions at energy $E$ in the interior and match them to real Coulomb functions at $R_{\mathrm{box}}$ (Sec.~\ref{subsec:Matching}). The complex incoming-wave combinations are then formed from these matched solutions (Sec.~\ref{subsec:Observables}).

Let $Q$ denote the space of basis functions that vanish at $R_{\mathrm{box}}$, and $P$ the surface space of the $M$ open channels at energy $E$. Closed-channel components remain coupled to the open channels in $Q$, with their surface coefficients set to zero. Diagonalization in $Q$ gives the \emph{box states} $|\Phi_i\rangle$, $i=1,\ldots,N$,
\begin{equation}\label{eq:BoxStates}
\langle\Phi_i|\hat{H}|\Phi_j\rangle=E_i\,\delta_{ij},\qquad
\langle\Phi_i|\Phi_j\rangle=\delta_{ij}.
\end{equation}
Away from the box eigenvalues, a set of linearly independent scattering solutions at energy E, each labeled by a different open channel $\beta$, is
\begin{equation}\label{eq:Standard_Scatt}
|\Psi_{\beta E}\rangle = |B_\beta\rangle+
\sum_{i=1}^{N} |\Phi_i\rangle\,
\frac{\langle\Phi_i|(E-\hat{H})|B_\beta\rangle}{E_i-E}.
\end{equation}
Equation~\eqref{eq:Standard_Scatt} satisfies the Schr\"odinger equation projected on $Q$, i.e., the Galerkin conditions $\langle\Phi_i|(\hat H-E)|\Psi_{\beta E}\rangle=0$ for every $i$ in $Q$. Evidently, missing radial, angular, and correlation components can affect both the interior wave function and the matched observables. For notational simplicity in the following, we define the $N\times M$ surface-coupling and interior coefficient matrices $\mathsf h$ and $X$ through their elements,
\begin{equation}\label{eq:SurfaceCouplings}
h_{i\beta}(E)\equiv\langle\Phi_i|(\hat{H}-E)|B_\beta\rangle,
\qquad
X_{i\beta}(E)=\frac{h_{i\beta}(E)}{E-E_i},
\end{equation}
with which
\begin{equation}\label{eq:SpectralCoefficientExpansion}
|\Psi_{\beta E}\rangle = |B_\beta\rangle + \sum_{i=1}^N |\Phi_i\rangle X_{i\beta}(E)
\end{equation}
The box eigenpairs and the Hamiltonian and overlap matrix elements connecting them to the surface functions are computed once for each symmetry block. Their combination $\mathsf h(E)$ is assembled at each scattering energy. 

As already noted in~\cite{Borras2024}, Equation~\eqref{eq:Standard_Scatt} is poorly conditioned near a box eigenvalue $E_j$: its columns contain large, nearly parallel contributions proportional to $(E-E_j)^{-1}$, whose cancellation is required to recover the finite physical combinations in the asymptotic match. At $E=E_j$ the expression is undefined. These are coordinate singularities of the auxiliary representation, however, not singularities of the scattering problem, and hence they can be removed. The method described in~\cite{Borras2024} to do so is applicable only to the case in which each discretized-continuum eigenvalue of the Hamiltonian projected onto Q is non-degenerate. Apart for accidental degeneracies, this is generally the case if Q correspond to a well defined component of a (possibly multidimensional) irreducible representation of the Hamiltonian's full invariance space group. When the invariance group is non abelian but one only classifies the states according to the irreducible representations of a subgroup of D2h, which is frequently the case for molecular configuration-interaction codes, some eigenvalues may be degenerate because they correspond to multiple components of a same irreducible representation of the actual invariance group of the system. In the following, we describe a procedure that applies to the degenerate case, and more generally to the case of several closely-clustered eigenvalues.

At any energy $E$ above the first ionization threshold $E_{\mathrm{th}}$, let \(E_j\) be the nearest box eigenvalue above that threshold. We define \(\mathcal J(E)=\{j_1,\ldots,j_g\}\) as the set of indices of all box eigenvalues above threshold within a tolerance $\tau(E)$ of \(E_j\).
We then write the ansatz for the $M$ auxiliary solutions as
\begin{equation}\label{eq:PoleFreeStates}
\begin{split}
|\widetilde\Psi_{bE}\rangle={}&
\sum_{k=1}^{g}|\Phi_{j_k}\rangle\,C^{\mathcal J}_{kb}
+\sum_{i\notin\mathcal J}|\Phi_i\rangle\,
 \frac{[\mathsf h\,\mathsf L]_{ib}}{E-E_i}+\sum_\alpha|B_\alpha\rangle\,\mathsf L_{\alpha b}.
\end{split}
\end{equation}
where $C^{\mathcal{J}}$, a $g\times M$ matrix, and $\mathsf{L}$, an $M\times M$ matrix, are unknowns, and the notation $[\mathsf{h}\mathsf{L}]_{ib}$ stands for $\sum_\alpha h_{i\alpha} L_{\alpha b}$.
Let's define
\begin{equation}\label{eq:PoleCluster}
\begin{aligned}
D^{\mathcal J}&=
\operatorname{diag}_{k=1,\ldots,g}(E-E_{j_k}),\\
H^{\mathcal J}_{k\alpha}&=h_{j_k\alpha}(E).
\end{aligned}
\end{equation}
The condition for the auxiliary functions $|\tilde{\Psi}_{bE}\rangle$ to satisfy the projected Schrödinger equation is then
\begin{equation}
D^{\mathcal J}C^{\mathcal J} = H^{\mathcal J}\mathsf L.
\end{equation}
If we define the rectangular matrices
\begin{equation*}
\mathcal B^{\mathcal J}\equiv
\begin{bmatrix}D^{\mathcal J},&-H^{\mathcal J}\end{bmatrix},
\qquad
\mathsf Z\equiv\begin{bmatrix}C^{\mathcal J}\\ \mathsf L\end{bmatrix},
\end{equation*}
this equation can be written as
\begin{equation}\label{eq:PoleBoundaryTransform}
\mathcal B^{\mathcal J}(E)\mathsf Z=0.
\end{equation}
We determine the right null space of the $g\times(g+M)$ matrix $\mathcal B^{\mathcal J}$ through its full singular-value decomposition (SVD),
\begin{equation}\label{eq:BorderedSVD}
\mathcal B^{\mathcal J}
=U^{\mathcal J}\Sigma^{\mathcal J}(V^{\mathcal J})^\dagger.
\end{equation}
Here $U^{\mathcal J}$ and $V^{\mathcal J}$ are unitary matrices of dimensions $g\times g$ and $(g+M)\times(g+M)$, respectively. Their columns are the left and right singular vectors. The $g\times(g+M)$ matrix $\Sigma^{\mathcal J}$ has the form
\begin{equation}\label{eq:BorderedSingularValues}
\Sigma^{\mathcal J}=
\begin{bmatrix}
\operatorname{diag}(\sigma_1,\ldots,\sigma_g)&0_{g\times M}
\end{bmatrix},
\end{equation}
where the singular values are real and nonnegative, ordered as $\sigma_1\ge\cdots\ge\sigma_g\ge0$. If $r$ denotes the rank of $\mathcal B^{\mathcal J}$, the first $r$ singular values are nonzero, and columns $r+1,\ldots,g+M$ of $V^{\mathcal J}$ span its right null space. For full row rank, $r=g$, the last $M$ columns of $V^{\mathcal J}$ form $\mathsf Z$: their first $g$ rows give $C^{\mathcal J}$ and their remaining $M$ rows give $\mathsf L$. If $r<g$, the right null space contains $g-r$ additional directions associated with decoupled bound states. We remove these directions as described below and partition the remaining $M$ scattering columns in the same way. This is the central pole-removal step: none of the denominators belonging to $\mathcal J$ is ever formed. The complete interior coefficient matrix has elements
\begin{equation}\label{eq:PoleFreeCoefficients}
\widetilde C_{ib}=
\begin{cases}
C^{\mathcal J}_{kb},&i=j_k\in\mathcal J,\\[1mm]
\displaystyle \frac{[\mathsf h\,\mathsf L]_{ib}}{E-E_i},&i\notin\mathcal J.
\end{cases}
\end{equation}
This construction avoids division by the energy differences associated with every level in \(\mathcal J\). For a single selected level, it removes the corresponding pole; for several selected levels, it treats their contributions jointly while keeping track of their individual eigenvalues.

The additional decoupled states are identified through the left null space, spanned by columns $r+1,\ldots,g$ of $U^{\mathcal J}$ when $r<g$. Its vectors $u$ satisfy
\begin{equation}\label{eq:DecoupledBoxStates}
D^{\mathcal J}u=0,\qquad (H^{\mathcal J})^\dagger u=0.
\end{equation}
Each such vector defines a state $|\chi_u\rangle=\sum_{k=1}^{g}u_k|\Phi_{j_k}\rangle$ with zero boundary values that satisfies the variational projected eigenvalue equation in both $Q$ and $P$ (Sec.~\ref{subsubsec:VariationalMatching}). It is therefore a bound eigenstate of the Hamiltonian in the chosen close-coupling model space, at an energy embedded in the open-channel continuum. Such decoupling can follow from an exact symmetry that the symmetry subgroup used in the computation does not discriminate. The boundary amplitudes do not fix the coefficients of these bound states in the auxiliary solutions. \ASTRA{} identifies these vectors through the left null space of $\mathcal B^{\mathcal J}$ and selects the orthogonal complement of the vectors $[u;0]$ in its right null space, leaving $M$ scattering columns. This sets the otherwise arbitrary coefficients of the decoupled states to zero and applies without changing any of the selected eigenvalues.

Any loss of row rank is determined from the rows of $H^{\mathcal J}$ corresponding to box eigenvalues $E_{j_k}=E$; every nonzero $E-E_{j_k}$ supplies an independent diagonal pivot in $\mathcal B^{\mathcal J}$. The rank of this coupling submatrix is evaluated by a separate, globally scaled SVD, with a threshold equal to machine precision times its largest singular value and its larger dimension. Small resolved singular values are kept: they may describe an extremely narrow resonance. A singular value below the rank threshold is treated as zero, even if this choice does not categorically exclude the possibility that the state in question does belong to the continuum. Indeed, while numerical symmetry breaking can give a symmetry-decoupled bound state a spurious width, a physical coupling can legitimately fall below numerical resolution. Distinguishing these cases, if necessary at all, requires additional symmetry and convergence checks.

\subsection{Variational surface matching\label{subsec:Matching}}

Beyond the matching radius $R_{\mathrm{box}}$, the reduced radial functions in the $M$ open channels are approximated by combinations of energy-normalized regular and irregular Coulomb functions, $\mathcal F_\beta(r)$ and $\mathcal G_\beta(r)$~\cite{Thompson1985,Seaton1983}. Closed channels remain in the box expansion with zero surface coefficients, as specified in Sec.~\ref{subsec:ScatteringStates}.
Diagonal Coulomb matching neglects anisotropic permanent-multipole and polarization couplings outside $R_{\mathrm{box}}$. For molecular targets, convergence with respect to the box radius $R_{\mathrm{box}}$ must therefore be checked for phase-sensitive observables. Within this outer-region approximation,
\begin{equation}\label{eq:AsymptoticExpansion}
\Psi_{bE}\underset{r\to\infty}{\longrightarrow}
\sum_\beta^{\mathrm{open}}\Upsilon_\beta
\left[\mathcal F_\beta(r)A_{\beta b}
+\mathcal G_\beta(r)B_{\beta b}\right],
\end{equation}
with $\mathbf A(E)$ and $\mathbf B(E)$ fixed by the boundary values and derivatives. We henceforth omit the tilde on the pole-free auxiliary states of Eq.~\eqref{eq:PoleFreeStates}, using $\widetilde C$ for their interior coefficients. With $v=B(R_{\mathrm{box}})$, only the surface functions contribute to the value table,
\begin{equation}\label{eq:ValueTable}
\mathcal V_{\beta b}\equiv
\big[\Psi_{bE}\big]_\beta(R_{\mathrm{box}})
=v\,\mathsf L_{\beta b}.
\end{equation}
For the original auxiliary solutions of Eq.~\eqref{eq:Standard_Scatt}, $\mathsf L=\mathbb 1$ and $\mathcal V=v\mathbb 1$. The values are thus fixed by the chosen auxiliary basis; the substantive choice in the matching is how to evaluate the channel slopes,

\begin{equation}\label{eq:DerivativeTableDef}
\mathcal{T}_{\beta b}\equiv\frac{d}{dr}\big[\Psi_{bE}\big]_\beta\Big|_{R_{\mathrm{box}}}.
\end{equation}
The Coulomb normalization factor $\sqrt{2/\pi k_\beta}$ gives the channel-independent Wronskian
\begin{equation}\label{eq:Wronskian}
W = \mathcal{F}_\beta'\mathcal{G}_\beta-\mathcal{F}_\beta\mathcal{G}_\beta' =
\frac{2}{\pi}.
\end{equation}
The exact $2\times2$ match to $\mathcal V$ and $\mathcal T$ in each channel, requiring no fit, then gives
\begin{equation}\label{eq:MatchingSystem}
\mathbf{A} =
W^{-1}\big(\mathcal{G}\,\mathcal{T}-\mathcal{G}'\,\mathcal{V}\big),
\quad
\mathbf{B} =
W^{-1}\big(\mathcal{F}'\,\mathcal{V}-\mathcal{F}\,\mathcal{T}\big),
\end{equation}
with $\mathcal{F}$, $\mathcal{G}$, $\mathcal{F}'$, $\mathcal{G}'$ the diagonal matrices of boundary values of the Coulomb pair. The reaction and scattering matrices are~\cite{Newton}
\begin{equation}\label{eq:KMatrix}
\begin{aligned}
\mathbf K&=\mathbf B\mathbf A^{-1}
  =\mathbf O\tan(\bm\delta)\mathbf O^\dagger,\\
\mathbf S&=\frac{\mathbb 1+i\mathbf K}{\mathbb 1-i\mathbf K}=\mathbf O\,e^{2i\bm\delta}\,\mathbf O^\dagger,
\end{aligned}
\end{equation}
where, for Hermitian $\mathbf K$, $\mathbf O$ is unitary, its columns are the orthonormal eigenchannels, and $\bm\delta$ is the diagonal matrix of eigenphase shifts. In the absence of scattering relative to the Coulomb reference, $\mathbf B=0$, $\mathbf K=0$, and $\mathbf S=\mathbb 1$. At a coordinate pole of $\mathbf K$, the physical $\mathbf S$ can be evaluated directly from $\mathbf A$ and $\mathbf B$ through Eq.~\eqref{eq:SMatrixFromAB}, without inverting $\mathbf A$.
In the real, spin-free representation used for the present calculations, $\mathbf K=\mathbf K^{t}$ and $\mathbf S=\mathbf S^{t}$.

\subsubsection{Direct Coulomb matching and its residual}

With $v'=B'(R_{\mathrm{box}})$, direct differentiation of the interior expansion, as used in the earlier construction, gives

\begin{equation}\label{eq:DirectCoulombDerivative}
\mathcal T=v'\,\mathsf L+D\,\widetilde C,
\qquad
D_{\beta i}=\frac{d}{dr}\big[\Phi_i\big]_\beta\Big|_{R_{\mathrm{box}}},
\end{equation}
where $D_{\beta i}$ is determined by the penultimate B-spline coefficient of $\Phi_i$ in channel $\beta$. We refer to the use of this derivative in Eq.~\eqref{eq:MatchingSystem} as \emph{direct Coulomb matching}. This is also the prescription described for \XCHEM{}-2.0: its scattering solutions are matched to Coulomb functions using their boundary values and radial derivatives [Sec.~2.3, Eq.~(18) of Ref.~\cite{Borras2024}].

Direct differentiation is linear in $\widetilde C$, so an interior coefficient error $\delta\widetilde C$ propagates to the slope at first order. In particular, the component $\eta$ of the exact solution omitted from the computational space produces an $O(\eta)$ error. Moreover, before transformation by $\mathsf L$, the channel couplings involve products $D_{\beta i}h_{i\alpha}$ that do not generally form a Hermitian matrix; neither Hermiticity of $\mathbf K$ nor unitarity of $\mathbf S$ is automatically enforced. The accuracy with which the flux is conserved, therefore, can be measured by $\|\mathbf S^\dagger\mathbf S-\mathbb 1\|$, where $\mathbf S$ is computed independently of $\mathbf K$ using Eq.~\eqref{eq:SMatrixFromAB}.

\subsubsection{A variational correction to the surface derivative\label{subsubsec:VariationalMatching}}

Thankfully, this linear dependence of the boundary condition on the error on the interior functions can be eliminated by replacing the directly evaluated slope by
\begin{equation}\label{eq:TableGap}
\begin{aligned}
\mathcal T^{\mathrm S}&=\mathcal T+\frac{2}{v}\,\rho,\\
\rho_{\beta b}&=\langle B_\beta|(\hat H-E)|\Psi_{bE}\rangle.
\end{aligned}
\end{equation}
The correction is the residual of the Schr\"odinger equation projected on the surface functions, which vanishes for an exact solution. For the Galerkin solutions, it cancels the first-order error of the direct derivative, as we now show.

Consider first the unit-boundary solutions of Eq.~\eqref{eq:Standard_Scatt}, away from a box eigenvalue. For generally complex interior functions $u$ and $w$, define the finite-volume sesquilinear form
\begin{equation}\label{eq:BoundaryBilinearForm}
\begin{aligned}
\mathfrak a_E[u,w]
&=\langle u|(\hat H-E)|w\rangle\\
&+\frac12\sum_\gamma u_\gamma^*(R_{\mathrm{box}})\,w'_\gamma(R_{\mathrm{box}}),
\end{aligned}
\end{equation}
where $u_\gamma$ and $w_\gamma$ are the reduced radial channel components. 
Integrating the radial kinetic-energy matrix element by parts gives a boundary contribution that cancels the explicit surface term. The remaining kinetic contribution to \(\mathfrak a_E[u,w]\) is \(\frac12\sum_\gamma\int_0^{R_{\mathrm{box}}}(u'_\gamma)^*w'_\gamma\,dr\). With Hermitian channel interactions, the form obeys $\mathfrak a_E[u,w]=\mathfrak a_E[w,u]^*$. 
Since the box states vanish at the boundary, the Galerkin equations give $\mathfrak a_E[\Phi_i,\Psi_{\alpha E}]=0$. We define the matrix of the sesquilinear form on the auxiliary solutions as
\begin{equation}\label{eq:BridgeIdentity}
\begin{aligned}
\mathcal R_{\beta\alpha}
&\equiv\mathfrak a_E[\Psi_{\beta E},\Psi_{\alpha E}]=\mathfrak a_E[B_\beta,\Psi_{\alpha E}] =\\
&=\rho_{\beta\alpha}+\frac v2\,\mathcal T_{\beta\alpha}
 =\frac v2\,\mathcal T^{\mathrm S}_{\beta\alpha}.
\end{aligned}
\end{equation}
The corrected derivative consequently evaluates a Hermitian sesquilinear form on the interior solutions.

To establish its stationarity, let $\Psi^{\mathrm{ex}}_{\alpha E}$ solve the same interior problem exactly, with the same boundary values, and write $\Psi_{\alpha E}=\Psi^{\mathrm{ex}}_{\alpha E}+\eta_\alpha$. The error $\eta_\alpha$ vanishes at the boundary. Hence $\mathfrak a_E[\eta_\beta,\Psi^{\mathrm{ex}}_{\alpha E}]=0$, and Hermiticity eliminates both cross terms in the difference
\begin{equation}\label{eq:ResponseError}
\mathcal R_{\beta\alpha}-\mathcal R^{\mathrm{ex}}_{\beta\alpha}
=\mathfrak a_E[\eta_\beta,\eta_\alpha].
\end{equation}
The error is quadratic in a norm for which the form is bounded. 
Therefore,
\begin{equation}\label{eq:OrderCounting}
\mathcal T^S-\mathcal T_{\mathrm{exact}}
=\frac{2}{v}
(\mathcal R-\mathcal R^{\mathrm{ex}}) = O(\|\eta\|^2),
\end{equation}
whereas $\mathcal T-\mathcal T_{\mathrm{exact}}$ keeps the term $-2\rho/v$, which is linear in the error in the interior wave function: the correction in Eq.~\eqref{eq:TableGap} removes the linear term. The difference between the two derivatives therefore estimates the leading error of direct differentiation. 
We can take advantage of this quadratic dependence by using $\mathcal{T}^S$ instead of $\mathcal{T}$ in \eqref{eq:MatchingSystem}.

Hermiticity of $\mathfrak a_E$ also gives $\mathcal V^\dagger\mathcal T^{\mathrm S}=(\mathcal T^{\mathrm S})^\dagger\mathcal V$, including in the pole-free auxiliary basis. With equal channel Wronskians, Eq.~\eqref{eq:MatchingSystem} then implies $\mathbf A^\dagger\mathbf B=\mathbf B^\dagger\mathbf A$, ensuring a Hermitian $\mathbf K$ and a unitary $\mathbf S$. The variational match thus conserves flux at any basis truncation.

\subsection{Incoming-wave states, scattering matrix, and photoionization
observables\label{subsec:Observables}}

Since, asymptotically,
\begin{equation}\label{eq:ExponentialCombinations}
\sqrt{\frac{2}{\pi k_\beta}}\, e^{\pm i\theta_\beta(r)} =\mathcal{G}_\beta(r)
\pm i \mathcal{F}_{\beta}(r),
\end{equation}
the interior solutions~\eqref{eq:AsymptoticExpansion} carry outgoing amplitude $\frac{1}{2}(\mathbf{B}-i\mathbf{A})$ and incoming amplitude $\frac{1}{2}(\mathbf{B}+i\mathbf{A})$ in each channel. Equation~\eqref{eq:ExponentialCombinations} shows that the following linear combination has exactly the channel-dependent prefactor $1/\sqrt{2\pi k_\beta}$ required by the incoming boundary condition~\eqref{eq:IncomingBC}:
\begin{equation}\label{eq:PsiMinusConstruction}
|\Psi_{\alpha E}^-\rangle
= \sum_b |\Psi_{bE}\rangle
\left[(\mathbf{B}-i\mathbf{A})^{-1}\right]_{b\alpha},
\end{equation}
whose outgoing amplitude is diagonal and whose incoming amplitude, on comparison with the relative minus in Eq.~\eqref{eq:IncomingBC}, gives
\begin{equation}\label{eq:SMatrixFromAB}
\begin{aligned}
\mathbf S
  &=-(\mathbf B-i\mathbf A)(\mathbf B+i\mathbf A)^{-1},
\end{aligned}
\end{equation}
consistently with the Cayley form implied by Eq.~\eqref{eq:KMatrix}. 
The one-photon observables follow from the dipole amplitudes $\langle\Psi^-_{\alpha E}|\hat\epsilon\cdot\vec d\,|g\rangle$ between the correlated initial state $|g\rangle$ and the incoming-wave states. \ASTRA{} evaluates them with the same TDM machinery used for the Hamiltonian~\cite{Randazzo2023}. For the spin-free photoionization considered here, matrix elements in the real-harmonic basis of Eq.~\eqref{eq:PWCDefinition} are transformed to the complex-harmonic combinations $Y_{\ell m}$ used below.

The total photoionization cross section can be computed by two independent numerical routes. The optical theorem~\cite{Newton} gives
\begin{equation}\label{eq:opticaltheorem}
\sigma_{\mathrm{tot}}(\omega)=-\frac{4\pi\omega}{c}\,\mathrm{Im}\,
\langle
g|\,d_\epsilon^\dagger\,G^+(E_g+\omega)
\,d_\epsilon\,|g\rangle,
\end{equation}
where $d_\epsilon=\hat\epsilon\cdot\vec d$ and $G^+(E)=(E-\hat H+i0^+)^{-1}$ is the retarded resolvent. 
We average this expression over three orthogonal Cartesian polarizations to compare with the isotropic cross sections below. In \ASTRA{}, a CAP emulates outgoing Siegert boundary conditions in the resolvent~\cite{Siegert1939,McCurdy1979,Randazzo2023}.
Alternatively, we sum the partial cross sections obtained from the explicit scattering states,
\begin{equation}\label{eq:SumPPCS}
\begin{split}
\sigma_{\mathrm{tot}}(\omega)&=\sum_a \sigma_{a}(\omega),\\
\sigma_a(\omega)&=\frac{4\pi^2\omega}{3c}\sum_{\ell m\tau}
\big|\langle\Psi_{a,E\ell m}^-|d_{1\tau}|g\rangle\big|^2,
\end{split}
\end{equation}
where $a$ labels the ionic state and $d_{1\tau}$ is a spherical dipole component; the average is over an isotropic initial ensemble (Appendix~\ref{app:OrientationalAverage}). The two routes share the Hamiltonian and electronic-structure ingredients but use different continuum boundary-condition solvers: an absorbed resolvent in one case and explicitly matched scattering states in the other. Their agreement, documented in Sec.~\ref{sec:Results}, is therefore a consistency check of the implementation.

The angularly resolved observables are derived in Appendix~\ref{app:OrientationalAverage}. In the molecular frame, the partial differential cross section for polarization $\hat\epsilon$ is given by Eq.~\eqref{eq:MFPADs}; for randomly oriented molecules, the laboratory-frame distribution takes the familiar form
\begin{equation}\label{eq:LFPADShape}
\frac{d\sigma_a^{\textsc{lf}}}{d\Omega}
=\frac{\sigma_a^{\textsc{lf}}}{4\pi}\Big[1+\beta_{1a}\,
P_1(\cos\theta) + \beta_{2a}\, P_2(\cos\theta)\Big],
\end{equation}
with $\beta_{1a}$ and $\beta_{2a}$ given by Eq.~\eqref{eq:BetaParameters}. For linear polarization the dipolar term $\beta_{1a}$ vanishes. In the electric-dipole approximation it can survive with circular polarization only in a chiral ensemble with nonzero enantiomeric excess~\cite{Ritchie1976,Goetz2016}. 

\subsection{Computational parameters}\label{Subsec:CompDetails}

The many-electron calculations follow the electronic-structure procedure of Ref.~\cite{Randazzo2023}. We obtain Hartree--Fock orbitals with \textsc{dalton}~\cite{Dalton2014} and optimize the molecular orbitals further by complete-active-space self-consistent-field (CASSCF) calculations with \textsc{lucia}~\cite{Olsen1988,Olsen2000}, which also supplies the ionic CASCI states and their TDMs. CAS($n$,$m$) denotes $n$ electrons in $m$ active orbitals. GBTOlib~\cite{Masin2020,GBTOlib-4} evaluates the one- and two-electron integrals in the hybrid Gaussian--B-spline basis. We remove linear dependencies with a deletion threshold of $10^{-5}$ and use B-splines of order 7, continued from the molecular region to $R_{\mathrm{box}}$. Table~\ref{tab:CompParams} collects the parameters of the photoionization calculations; the one-electron $\mathrm H_2^+$ calculation is specified in Sec.~\ref{Subsec:H2plusResults}.

\begin{table}[tb]
\caption{\label{tab:CompParams}Parameters of the many-electron photoionization calculations. Ionic-state counts refer to components in the Abelian computational group. All listed ionic states have full radial channels. $N_B$ and $\Delta r$ specify the order-7 B-spline basis within $R_{\mathrm{mol}}$; the grid continues to $R_{\mathrm{box}}$. The continuum and potential expansions terminate at $\ell_{\mathrm{max}}$ and $\lambda_{\mathrm{max}}$, respectively.}
\begin{ruledtabular}
\begin{tabular}{lccc}
Parameter & Ne & CO & pyrazine \\
\colrule
Gaussian basis & aug-cc-pVQZ & cc-pVTZ & cc-pVTZ \\
Orbital optimization & HF & CASSCF & CASSCF \\
Parent-ion CI & CAS(7,4) & CAS(9,8) & CAS(9,13) \\
Continuum ionic states & 4 & 4 & 4 \\
$\ell_{\mathrm{max}}$ & 2 & 4 & 4 \\
$\lambda_{\mathrm{max}}$ & 3 & 5 & 5 \\
$R_{\mathrm{mol}}$ ($a_0$) & 20 & 20 & 15 \\
$N_B$ & 56 & 50 & 39 \\
$\Delta r$ ($a_0$) & 0.40 & 0.45 & 0.45 \\
$R_{\mathrm{box}}$ ($a_0$) & 400 & 200 & 200 \\
$R_{\textsc{cap}}$ ($a_0$) & 300 & 150 & 150 \\
\end{tabular}
\end{ruledtabular}
\end{table}

The initial state $|g\rangle$ is the lowest eigenstate of the close-coupling Hamiltonian in the totally symmetric singlet space. One-photon absorption reaches $^1\mathrm B_{1u}$, $^1\mathrm B_{2u}$, and $^1\mathrm B_{3u}$ in $D_{2h}$, or $^1\mathrm A_1$, $^1\mathrm B_1$, and $^1\mathrm B_2$ in $C_{2v}$. The optical-theorem calculation uses the same radial box with two absorbers of the form $-i\eta\,\theta(r-R_{\textsc{cap}})(r-R_{\textsc{cap}})^2$~\cite{Randazzo2023}: a weak absorber with $\eta=10^{-6}$~a.u. starts at $100\,a_0$, and a stronger one with $\eta=10^{-4}$~a.u. starts at the radius listed in Table~\ref{tab:CompParams}.

For neon, the aug-cc-pVQZ basis~\cite{Kendall1992} and the neutral Hartree--Fock $2s$ and $2p$ orbitals give single-configuration $2s^22p^5\;{}^2P^o$ and $2s2p^6\;{}^2S^e$ parent ions. The continuum includes the dipole-accessible partial waves through $\ell=2$. This single-reference model reproduces the CC ansatz used in the NewStock comparison.

For CO, we use cc-pVTZ~\cite{Dunning1989} at the equilibrium bond length $R_{\mathrm{CO}}=1.1283$~\AA{}, with C at $(0,0,0.6447)$~\AA{} and O at $(0,0,-0.4837)$~\AA{}. The calculation uses the $C_{2v}$ subgroup of $C_{\infty v}$. A neutral ground-state CASSCF calculation optimizes ten electrons in the eight valence orbitals $3\sigma$--$6\sigma$, $1\pi$, and $2\pi$, with the C and O $1s$ orbitals kept doubly occupied. CAS(9,8) CI in these orbitals supplies the four components of the $\mathrm X\,^2\Sigma^+$, $\mathrm A\,^2\Pi$, and $\mathrm B\,^2\Sigma^+$ ionic terms, all coupled to continuum partial waves through $\ell=4$.

For pyrazine, we use cc-pVTZ in $D_{2h}$ at the geometry of the \XCHEM{} comparison: N at $(0,0,\pm2.6714)$, C at $(0,\pm2.1361,\pm1.3157)$, and H at $(0,\pm3.8951,\pm2.3599)$, in bohr. The molecular plane is $yz$. A neutral ground-state CASSCF calculation places ten electrons in thirteen active orbitals, comprising the $\pi$/$\pi^*$ system and nitrogen lone pairs, augmented by low-lying $\sigma^*$/Rydberg-like orbitals; the remaining doubly occupied orbitals are inactive. In the close-coupling expansion, we include the four lowest CAS(9,13) ions, $^2\mathrm B_{1g}$, $^2\mathrm B_{2g}$, $^2\mathrm A_g$, and $^2\mathrm B_{1u}$, each with continuum partial waves through $\ell=4$. The published \XCHEM{}-2.0 application~\cite{Borras2024} uses a five-ion SA-CASSCF(9,8) model.

\section{Results\label{sec:Results}}

We first test the scattering construction against the exactly separable one-electron problem of $\mathrm H_2^+$ and then compare variational and direct matching within the same CO calculation. The subsequent comparisons with NewStock and \XCHEM{} test the many-electron implementation through independent predictions of channel-resolved photoionization observables.

\subsection{Accuracy for the two-center Coulomb problem}\label{Subsec:H2plusResults}

The hydrogen molecular ion provides a direct test of the scattering method in a polycentric potential, without electron correlation or a many-electron close-coupling approximation. At fixed internuclear separation, its Schr\"odinger equation separates exactly in prolate spheroidal coordinates~\cite{Bates1953,Power1973,Singor2023}. We compute the reference eigenphases by solving the separated angular and radial equations and express the resulting reaction matrix in the spherical channels used by \ASTRA{}. Appendix~\ref{app:H2plusBenchmark} gives the construction. The reference therefore bypasses the Gaussian--B-spline representation and the surface-matching procedure whose accuracy we wish to establish.

We use an internuclear distance $R=1.99933\,a_0$, aug-cc-pVQZ Gaussians, seventh-order B-splines with spacing $1/3\,a_0$, and a matching radius $R_{\mathrm{box}}=100\,a_0$. The calculations include the $^2\mathrm A_g$, $^2\mathrm B_{1u}$, $^2\mathrm B_{2u}$, and $^2\mathrm B_{3u}$ continuum symmetries, with partial waves through $\ell_{\mathrm{max}}=2$ or 4. Although there is only one residual-ion state, the two-center potential couples spherical partial waves: at $\ell_{\mathrm{max}}=4$, for example, the six channels of $^2\mathrm A_g$ form $|m|=0,2,4$ blocks of dimensions $3+2+1$. Thus the comparison tests multichannel surface matching as well as the radial representation.

\begin{table}[t]
\caption{\label{tab:H2plusBenchmark}%
Largest absolute differences between the \ASTRA{} eigenphases and the prolate-spheroidal numerical reference, for $\mathrm H_2^+$ at $R=1.99933\,a_0$, over the sampled energies in the indicated intervals. For $\ell_{\mathrm{max}}=2$, the $^2\mathrm B_{2u}$ interval ends below the pseudoresonance discussed in the text. The degenerate $^2\mathrm B_{3u}$ block agrees with $^2\mathrm B_{2u}$ to machine precision and is omitted.}
\begin{ruledtabular}
\begin{tabular}{lcc}
Symmetry & max. $|\Delta\delta|$ (rad)
  & $\varepsilon$ range (hartree) \\
\colrule
\multicolumn{3}{l}{$\ell_{\mathrm{max}}=2$:}\\
$^2\mathrm{A}_{g}$ & $1.9\times10^{-2}$ & $0.1\leq\varepsilon\leq1.2$ \\
$^2\mathrm{B}_{1u}$ & $2.9\times10^{-3}$ & $0.1\leq\varepsilon\leq1.2$ \\
$^2\mathrm{B}_{2u}$ & $5.1\times10^{-2}$ & $0.1\leq\varepsilon\leq0.55$ \\
\multicolumn{3}{l}{$\ell_{\mathrm{max}}=4$:}\\
$^2\mathrm{A}_{g}$ & $1.2\times10^{-2}$ & $0.1\leq\varepsilon\leq1.2$ \\
$^2\mathrm{B}_{1u}$ & $1.2\times10^{-2}$ & $0.1\leq\varepsilon\leq1.2$ \\
$^2\mathrm{B}_{2u}$ & $1.1\times10^{-2}$ & $0.1\leq\varepsilon\leq1.2$ \\
\end{tabular}
\end{ruledtabular}
\end{table}

With $\ell_{\mathrm{max}}=4$, the lower partial-wave eigenchannels agree with the reference to $10^{-4}$--$10^{-3}$~rad over the energy range considered. Table~\ref{tab:H2plusBenchmark} reports the largest differences in each symmetry, which reach $1.1$--$1.2\times10^{-2}$~rad. These maxima belong to eigenchannels dominated by the highest partial wave of their parity block, $f\sigma_u$, $f\pi_u$, $g\sigma_g$, and $g\delta_g$, for which the next coupled partial wave lies outside the expansion. The comparison with the separable solution resolves the angular-cutoff error while establishing the accuracy already attained in the lower channels. The degenerate $^2\mathrm B_{2u}$ and $^2\mathrm B_{3u}$ results agree to machine precision.

Increasing $\ell_{\mathrm{max}}$ from 4 to 8 reduces the selected lower-channel differences, by about a factor of two at intermediate energies for $p\sigma_u$ and by up to one to two orders of magnitude toward the upper end of the energy interval. Subsequent replacement of aug-cc-pVQZ by aug-cc-pV5Z, halving of the spline spacing, and doubling of $R_{\mathrm{box}}$ reduce the $p\sigma_u$ difference by further factors of approximately three, two, and two, respectively. The $s\sigma_g$ differences are already below the few-$10^{-4}$-rad accuracy resolved by the numerical reference for that channel. These refinements demonstrate convergence of the hybrid-basis scattering solution toward the separable result, independently of the representation of a many-electron target.

At the smaller angular cutoff, $\ell_{\mathrm{max}}=2$, the $^2\mathrm B_{2u}$ eigenphase exhibits a narrow $\pi$ rise near total energy $E=1.145$~hartree. The feature is absent from the separable reference and disappears when $\ell_{\mathrm{max}}$ is raised to 4. The exact comparison therefore also identifies this feature as a pseudoresonance of the truncated angular expansion.

\subsection{Stability of variational surface matching}\label{Subsec:MatchingComparison}

To isolate the benefit of the variational correction, we evaluate direct and variational matching for the same CO Hamiltonian and progressively coarsen the radial basis. The production spacing $\Delta r=0.45\,a_0$ gives 50 B-splines in the molecular region, $R_{\mathrm{mol}}=20\,a_0$. Increasing the spacing to $0.65$, $0.83$, and $1.0\,a_0$ reduces this number to 37, 30, and 26, respectively; each grid continues to $R_{\mathrm{box}}=200\,a_0$. Both prescriptions are evaluated at 501 uniformly spaced energies between each pair of successive ionic thresholds and at 751 energies from the $\mathrm B\,^2\Sigma^+$ threshold to 59~eV above it, in each of the three dipole-accessible total symmetries.

\begin{widetable}
\caption{\label{tab:MatchingDiagnostics}Comparison of variational and direct surface matching for CO as the radial grid is coarsened. The flux defect is $\|\mathbf S^\dagger\mathbf S-\mathbb 1\|_{\max}$, with $\|X\|_{\max}=\max_{\alpha\beta}|X_{\alpha\beta}|$ and $\mathbf S$ evaluated directly from the Coulomb coefficients. The additional inverse-matrix diagnostics are $\mathrm{asym}(\mathbf K^{-1})=\|\mathbf K^{-1}-(\mathbf K^{-1})^t\|_{\max}/\|\mathbf K^{-1}\|_{\max}$ and the relative gap $\|(\mathbf K^{\mathrm S})^{-1}-(\mathbf K^{\mathrm D})^{-1}\|_{\max}/\|(\mathbf K^{\mathrm S})^{-1}\|_{\max}$, where S and D denote variational and direct matching (the code stores $\mathbf K^{-1}/\pi$ for historical reasons). Values are maxima over the uniform scans unless marked as medians; matrix diagnostics precede symmetrization. Above B, the matrix and phase medians include three total symmetries; phase differences are the largest one-to-one paired differences modulo $\pi$. Cross-section medians include the four ionic components. Dashes indicate unavailable direct-matching observables.}
\begin{ruledtabular}
\begin{tabular}{llcccc}
Diagnostic & Method or region & \multicolumn{4}{c}{$\Delta r$ ($a_0$)} \\
& & 0.45 & 0.65 & 0.83 & 1.0 \\
\colrule
Flux defect & variational
  & $3{\times}10^{-11}$ & $2{\times}10^{-11}$
  & $1{\times}10^{-11}$ & $1{\times}10^{-11}$ \\
Flux defect & direct
  & -- & $1{\times}10^{-5}$ & $5{\times}10^{-5}$
  & $7{\times}10^{-4}$ \\
$\mathrm{asym}(\mathbf K^{-1})$ & variational
  & $3{\times}10^{-11}$ & $9{\times}10^{-12}$
  & $3.5{\times}10^{-11}$ & $8{\times}10^{-12}$ \\
$\mathrm{asym}(\mathbf K^{-1})$ & direct
  & $4{\times}10^{-6}$ & $3{\times}10^{-5}$
  & $3{\times}10^{-4}$ & $2{\times}10^{-3}$ \\
\colrule
Relative $\mathbf K^{-1}$ gap & $\mathrm X$--$\mathrm A$
  & $6{\times}10^{-7}$ & $2{\times}10^{-6}$
  & $7{\times}10^{-6}$ & $2{\times}10^{-5}$ \\
Relative $\mathbf K^{-1}$ gap (median) & above $\mathrm B$
  & $1{\times}10^{-4}$ & $1{\times}10^{-3}$
  & $1{\times}10^{-2}$ & $4{\times}10^{-2}$ \\
$|\Delta\delta|$ (median) & above $\mathrm B$ (rad)
  & -- & $5.1{\times}10^{-5}$ & $3.6{\times}10^{-4}$
  & $1.5{\times}10^{-3}$ \\
$|\Delta\sigma_\alpha|/\sigma_\alpha$ (median) & above $\mathrm B$
  & -- & $2.3{\times}10^{-5}$ & $1.1{\times}10^{-4}$
  & $4.2{\times}10^{-4}$ \\
\end{tabular}
\end{ruledtabular}
\end{widetable}

The flux defect $\|\mathbf S^\dagger\mathbf S-\mathbb 1\|_{\max}$ gives the clearest comparison (Table~\ref{tab:MatchingDiagnostics}). For variational matching it remains between $10^{-11}$ and $3\times10^{-11}$ across all four bases. For direct matching it grows from $10^{-5}$ at $\Delta r=0.65\,a_0$ to $7\times10^{-4}$ at $\Delta r=1.0\,a_0$. The variational construction thus preserves flux conservation even when the radial basis is made substantially poorer. This numerical stability is the practical consequence of replacing the directly evaluated derivative by the stationary surface expression derived in Sec.~\ref{subsec:Matching}.

The contrast becomes more pronounced at higher energies, where more ionic channels are open. Between the first two thresholds, the maximum relative difference between the inverse reaction matrices grows from $6\times10^{-7}$ to $2\times10^{-5}$ over the four bases. Above the highest threshold, its median grows from $10^{-4}$ to $4\times10^{-2}$. In the coarsest basis, the prescriptions differ by a median of $1.5\times10^{-3}$~rad in the largest paired eigenphase difference and by $4.2\times10^{-4}$ in the partial cross sections. Variational matching maintains its flux-conservation accuracy throughout the coarsening of the basis, whereas the direct matching is uniformly orders of magnitude worse and rapidly deteriorates. Denser scans of the resonant region below the second threshold give a maximum variational flux defect of $7.8\times10^{-9}$, still well below the direct-matching defects in the coarser bases.

\subsection{Atomic ionization and degenerate box states: neon}\label{Subsec:Neon}

Neon tests a different aspect of the construction. NewStock exploits the full rotational symmetry of the atom, whereas \ASTRA{} classifies its states in the Abelian group $D_{2h}$. Multiple components of a rotational multiplet can then belong to the same computational symmetry and generate degenerate box eigenvalues. Such degeneracies are intrinsic to the atomic problem represented in this molecular basis. The comparison with NewStock consequently tests the treatment of degenerate eigenspaces in Sec.~\ref{subsec:ScatteringStates}, together with the angular coupling and normalization of the many-electron scattering amplitudes.

We use the single-reference ionic model of Sec.~\ref{Subsec:CompDetails} in both codes. Figure~\ref{Fig:Ne_OptTheor_Comparison} compares the total cross section and the $^2P^o$ asymmetry parameter on both sides of the $^2S^e$ threshold. The agreement extends from the smooth continuum into the narrow autoionizing structures below threshold. In particular, the two calculations reproduce the rapid changes of $\beta$ across these structures, while above threshold their asymmetry parameters agree to better than $10^{-3}$. The agreement of an angular observable, alongside the total yield, tests the relative partial-wave amplitudes and their interference through two different treatments of atomic symmetry.

\begin{widefigure}
\centering
\includegraphics[width=0.49\textwidth]{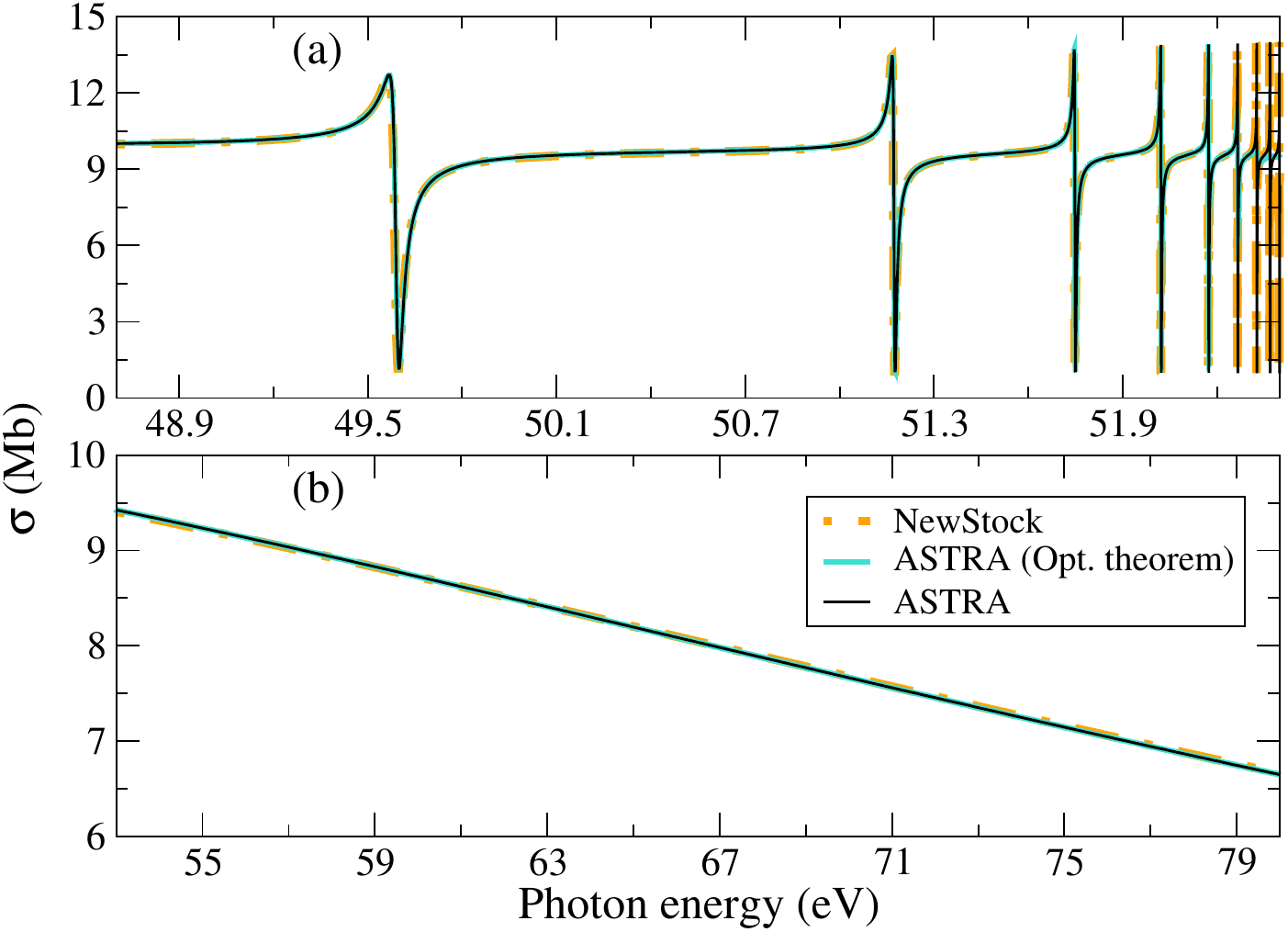}\hfill
\includegraphics[width=0.49\textwidth]{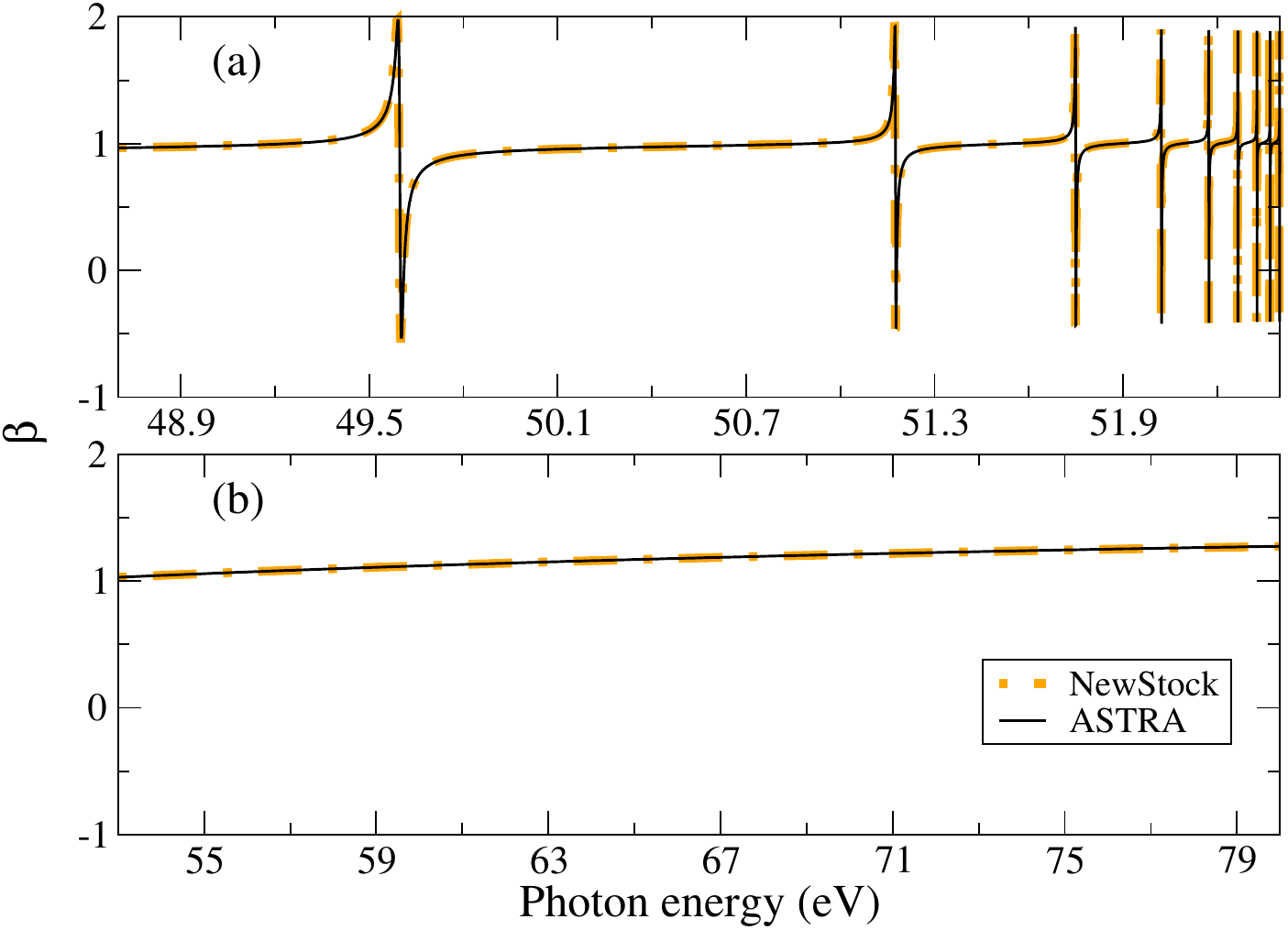}
\caption{\label{Fig:Ne_OptTheor_Comparison}\label{Fig:Ne_PartialBeta_SingleRef}Neon total photoionization cross section (left) and $^2P^o$ asymmetry parameter $\beta$ (right), below [panels (a)] and above [panels (b)] the $^2S^e$ threshold. NewStock is shown by dot-dashed curves and \ASTRA{} by solid curves, using single-reference parent ions in both codes. In the cross-section panels, the thick and thin solid curves are the \ASTRA{} optical-theorem and scattering-state results, respectively. The \ASTRA{} energy axis is shifted by $-0.075$~eV.}
\end{widefigure}

The partial cross sections provide a separate check of the division of the yield between the two ionic terms. Both the dominant $2p$ and the much weaker $2s$ channel agree with NewStock to median relative differences of approximately $0.2\%$ over the range extending to 285~eV. As an internal consistency check, the sum of the \ASTRA{} partial cross sections also reproduces the optical-theorem total: between 53 and 80~eV, the median and maximum relative differences are $4.1\times10^{-6}$ and $2.5\times10^{-5}$. Taken together, the independent atomic comparison and the internal sum check establish a consistent set of channel amplitudes despite the degeneracies inherited from the higher symmetry of the atom.

\subsection{Independent molecular calculations: carbon monoxide}\label{Subsec:CO}

For CO, we compare the \ASTRA{} predictions with those of the independent molecular ionization code \XCHEM{}~\cite{Marante2017,Borras2021,Borras2024}. Although both methods use a hybrid Gaussian--B-spline CC expansion, their quantum-chemistry support, one-electron bases, evaluation of integrals, and construction of scattering states are different. Agreement therefore tests the complete calculation of molecular ionization amplitudes through independent implementations. The comparison is particularly demanding for MFPADs, which preserves all those interferences between different ionization amplitudes that are lost in the LFPADs through the orientational averaging.

Figure~\ref{Fig:CO_CS_Comparison} compares the partial cross sections and laboratory-frame asymmetry parameters below and above the $\mathrm B\,^2\Sigma^+$ threshold. The calculations agree closely for the three open ionic terms above that threshold, with median relative cross-section differences of $0.73\%$, $0.18\%$, and $1.73\%$ for X, A, and B between 20.8 and 22.0~eV. The corresponding median absolute differences in $\beta$ are $0.013$, $0.0006$, and $0.021$. Below threshold, the agreement persists through the sequence of resonances in both partial yields and their angular asymmetries. The different signatures in the X and A channels are reproduced by both codes, providing a more discriminating comparison than agreement of their summed cross section alone.

\begin{widefigure}
\centering
\includegraphics[width=0.44\textwidth]{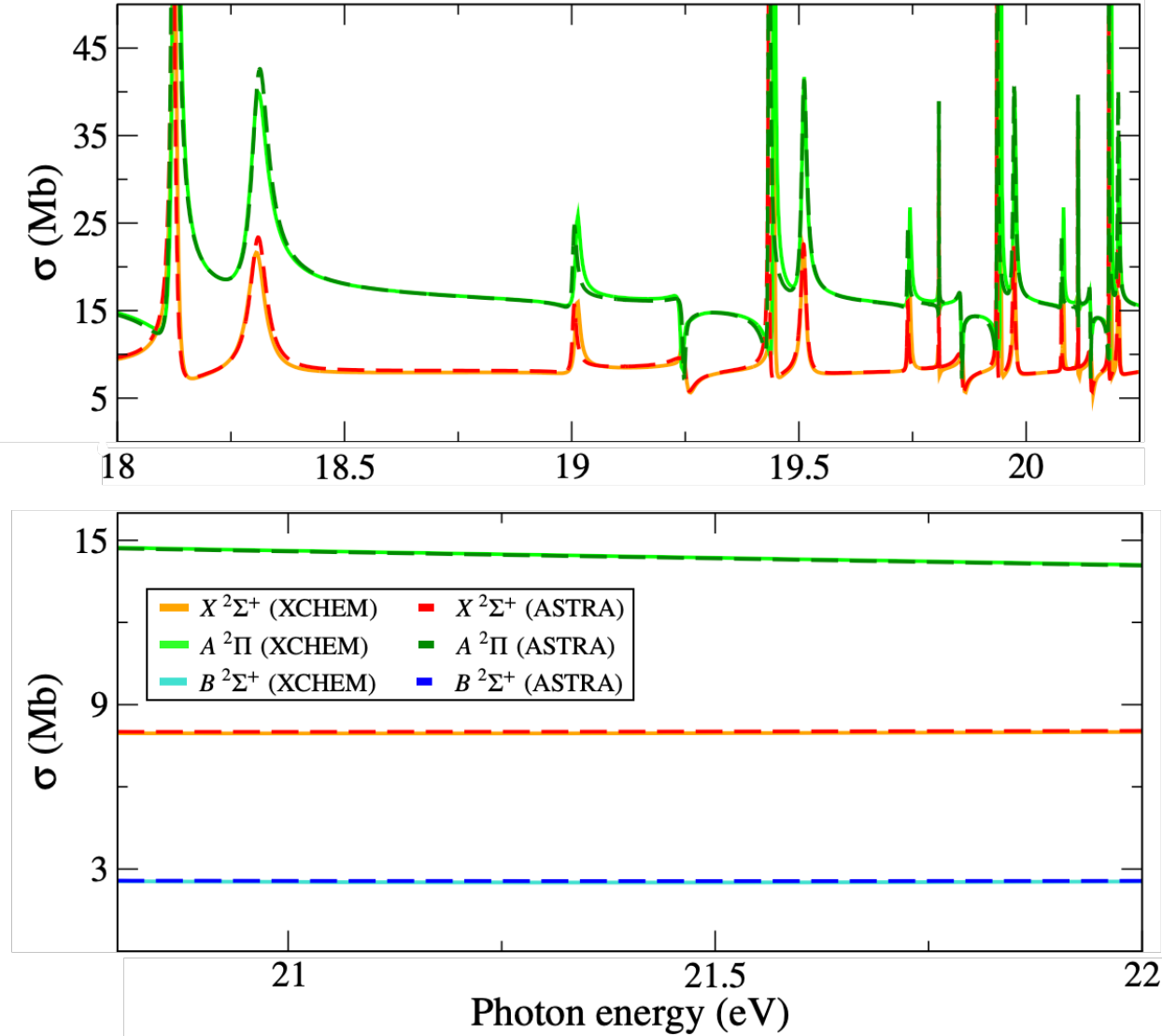}\hfill
\includegraphics[width=0.44\textwidth]{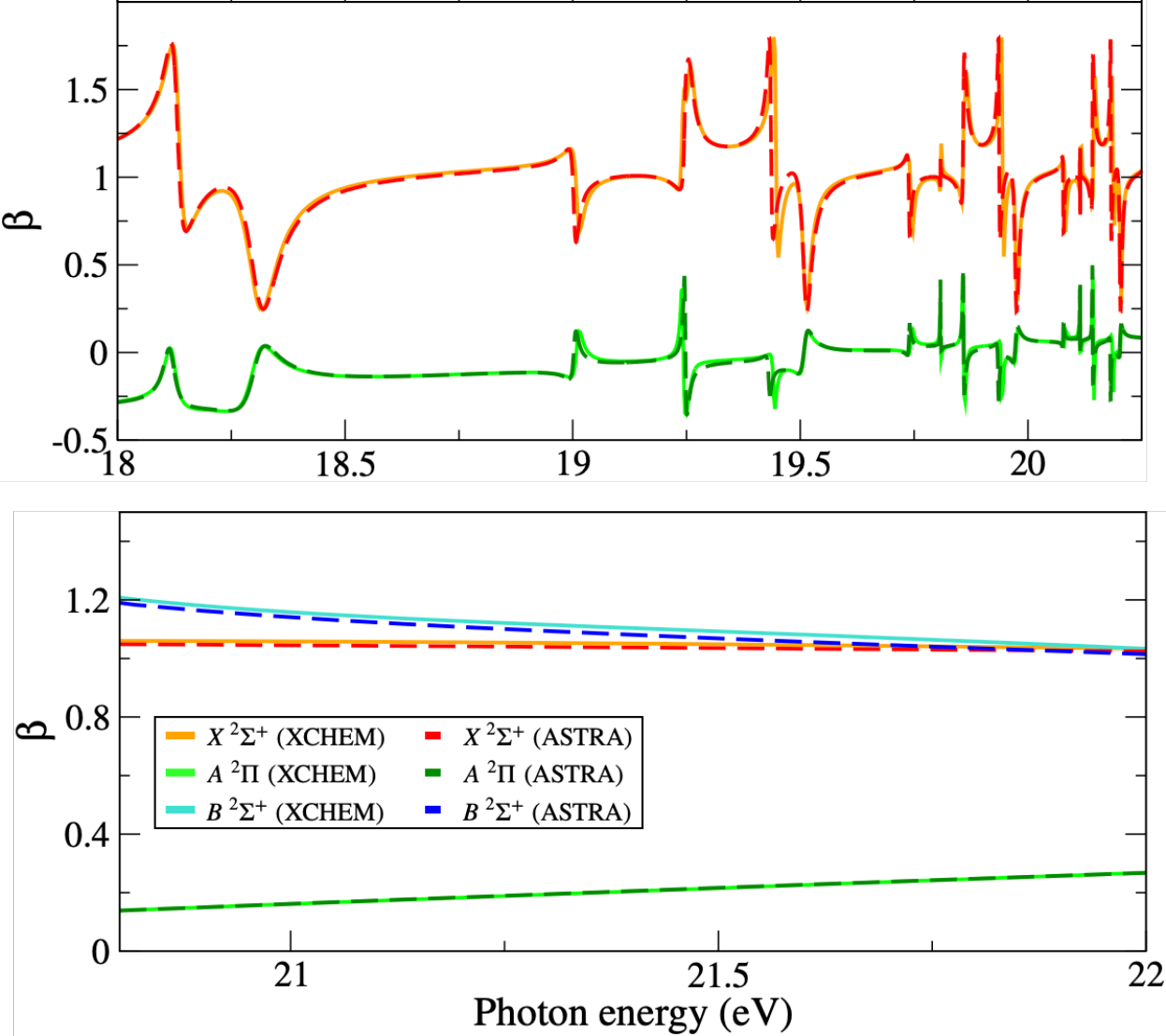}
\caption{\label{Fig:CO_CS_Comparison}\label{Fig:CO_Beta_Comparison}CO partial photoionization cross sections (left) and asymmetry parameters $\beta$ (right) from \ASTRA{} (dashed) and \XCHEM{} (solid). The upper panels lie between the $\mathrm A\,^2\Pi$ and $\mathrm B\,^2\Sigma^+$ thresholds; the lower panels lie above $\mathrm B\,^2\Sigma^+$. The same ionic channels are compared in the cross-section and angular-distribution panels.}
\end{widefigure}

The molecular-frame comparison brings the agreement into sharper focus. Figure~\ref{fig:CO_MFPAD1} shows the distributions for the $\mathrm X\,^2\Sigma^+$ and $\mathrm A\,^2\Pi_x$ ions with the light polarized along the molecular axis. At 18.75~eV the comparison samples the nonresonant background; the two higher energies, 18.99 and 19.01~eV, lie on opposite sides of a narrow autoionizing feature. Both codes reproduce the angular distributions in the background and their pronounced changes across the resonance, including the relative lobe strengths and their different evolution in the two ionic channels. The agreement therefore extends to the directional interference of the ionization amplitudes in the region where those amplitudes change most rapidly with energy.

\begin{widefigure}
\centering
\includegraphics[width=0.82\textwidth]{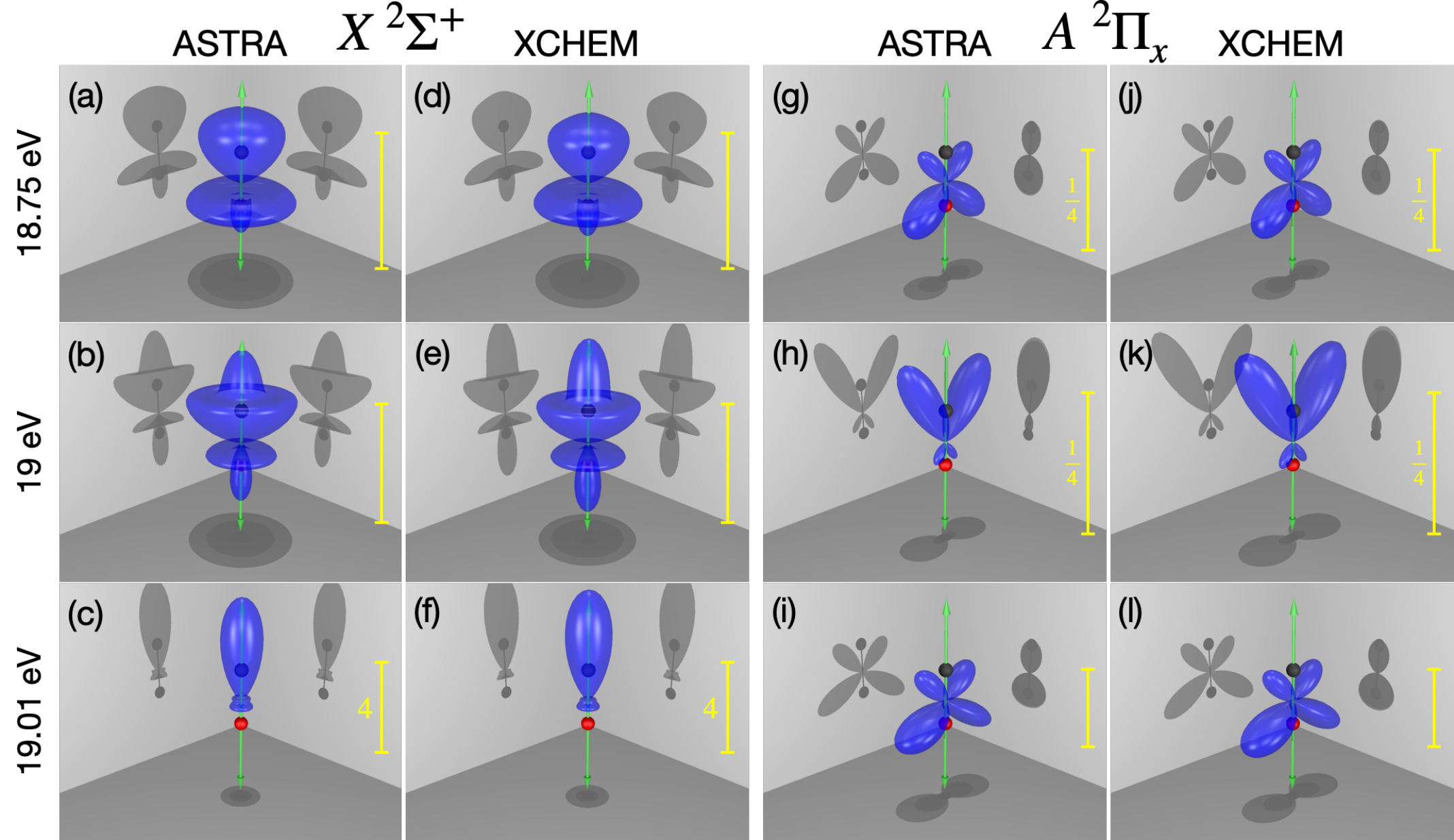}
\caption{\label{fig:CO_MFPAD1}CO molecular-frame angular distributions from \ASTRA{} and \XCHEM{} for the $\mathrm X\,^2\Sigma^+$ and $\mathrm A\,^2\Pi_x$ ions, at photon energies 18.75, 18.99, and 19.01~eV from top to bottom. The middle image label rounds 18.99 to 19~eV. The molecular axis and linear polarization lie along $z$, with C on the positive and O on the negative side. Each ion has adjacent \ASTRA{} and \XCHEM{} columns. Yellow segments give the radial scales, with the indicated multipliers applied separately to each panel. The radius represents the squared dipole angular amplitude $I_a(\hat k)$ in atomic units; multiplication by $4\pi^2\omega/c$ gives the differential cross section in atomic units of area per solid angle.}
\end{widefigure}

The planar cuts for perpendicular polarization in Fig.~\ref{fig:CO_MFPAD_Proj_Hor} provide a complementary comparison at 18.75~eV. The \ASTRA{} and \XCHEM{} curves closely overlap for both ionic channels in the two displayed planes, confirming the agreement for another polarization geometry. Here $\Pi_x$ labels one real component of the degenerate A ion. Agreement of the channel yields, laboratory-frame asymmetries, and molecular-frame distributions, both away from and across resonances, constitutes strong mutual validation of the two molecular ionization codes.

\begin{widefigure}
\centering
\includegraphics[width=\textwidth]{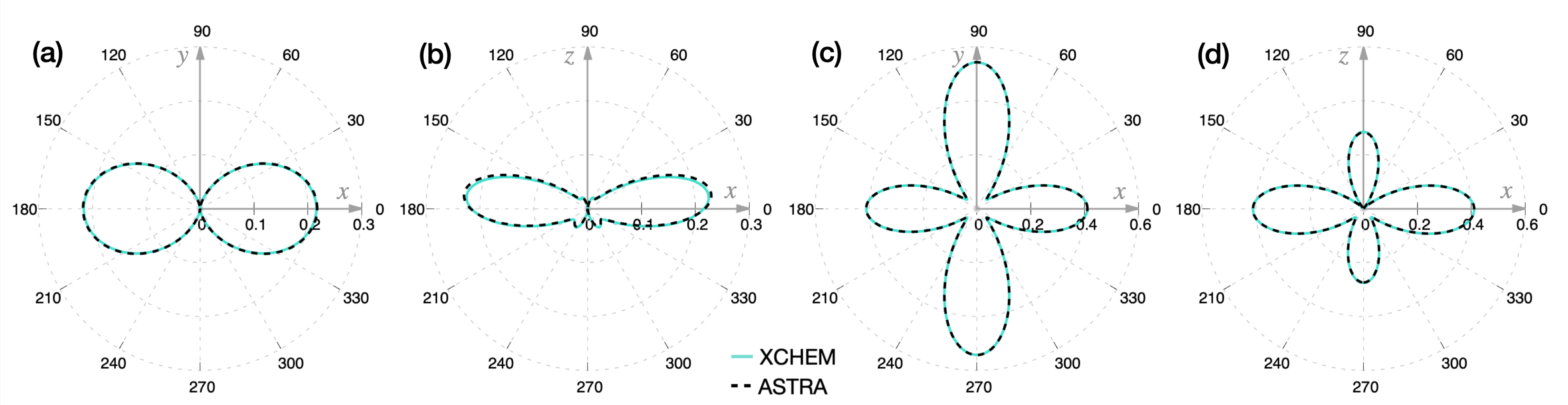}
\caption{\label{fig:CO_MFPAD_Proj_Hor}Cuts through the CO molecular-frame angular distributions in the $xy$ [(a), (c)] and $xz$ [(b), (d)] planes. \ASTRA{} (dashed) and \XCHEM{} (solid) are compared at 18.75~eV for light polarized along $\hat x$, leaving the ion in $\mathrm X\,^2\Sigma^+$ [(a), (b)] or $\mathrm A\,^2\Pi_x$ [(c), (d)]. The bond lies along $z$, with C at positive $z$. Each ionic-channel pair uses a common radial scale. The radius is $I_a(\hat k)$ in atomic units, as in Fig.~\ref{fig:CO_MFPAD1}.}
\end{widefigure}

\subsection{A larger polyatomic target: pyrazine}\label{Subsec:Pyrazine}

Pyrazine extends the independent comparison to a ten-atom, 42-electron molecule. This is an important test for the TDM formulation of \ASTRA{}, whose purpose is to make molecular CC calculations accessible with larger correlated ionic spaces. The \ASTRA{} close-coupling expansion includes the four lowest ions in CAS(9,13), as described in Sec.~\ref{Subsec:CompDetails}. The comparison concerns the partial cross sections in two intervals below the third ionization threshold.

Below the $^2\mathrm A_g$ threshold, only the $^2\mathrm B_{1g}$ ion is accessible, so its partial cross section equals the total [Fig.~\ref{Fig:Pyrazine_TotCrossSec}, left]. Above that threshold, both ionic channels contribute [right panels]. The calculations reproduce the overall magnitude and much of the dense resonance structure in both intervals, including the different strengths with which the structures appear in the two open channels. Differences in individual peak positions and heights are visible, and the agreement is less close than in CO. Nevertheless, the correspondence extends across a crowded spectrum and survives the opening of the second ionic channel, providing useful corroboration of both implementations in a substantially larger molecular calculation.

The remaining differences in individual resonance positions and profiles should also be considered in light of the different continuum representations. In \XCHEM{}, the inner region is described by atom-centered and diffuse monocentric Gaussian functions, while the B-splines begin at a radius chosen to make their overlap and exchange with the ionic orbitals negligible~\cite{Borras2024}. Increasing this radius places greater demands on the Gaussian representation of the continuum. For a molecule of the spatial extent of pyrazine, convergence with respect to the diffuse Gaussian basis and the B-spline onset radius is therefore an important consideration. \ASTRA{} instead includes B-splines throughout the molecular region and evaluates the mixed Gaussian--B-spline matrix elements. This difference provides a possible source of the residual discrepancies between the calculations.

\begin{widefigure}
\centering
\includegraphics[width=0.49\textwidth]{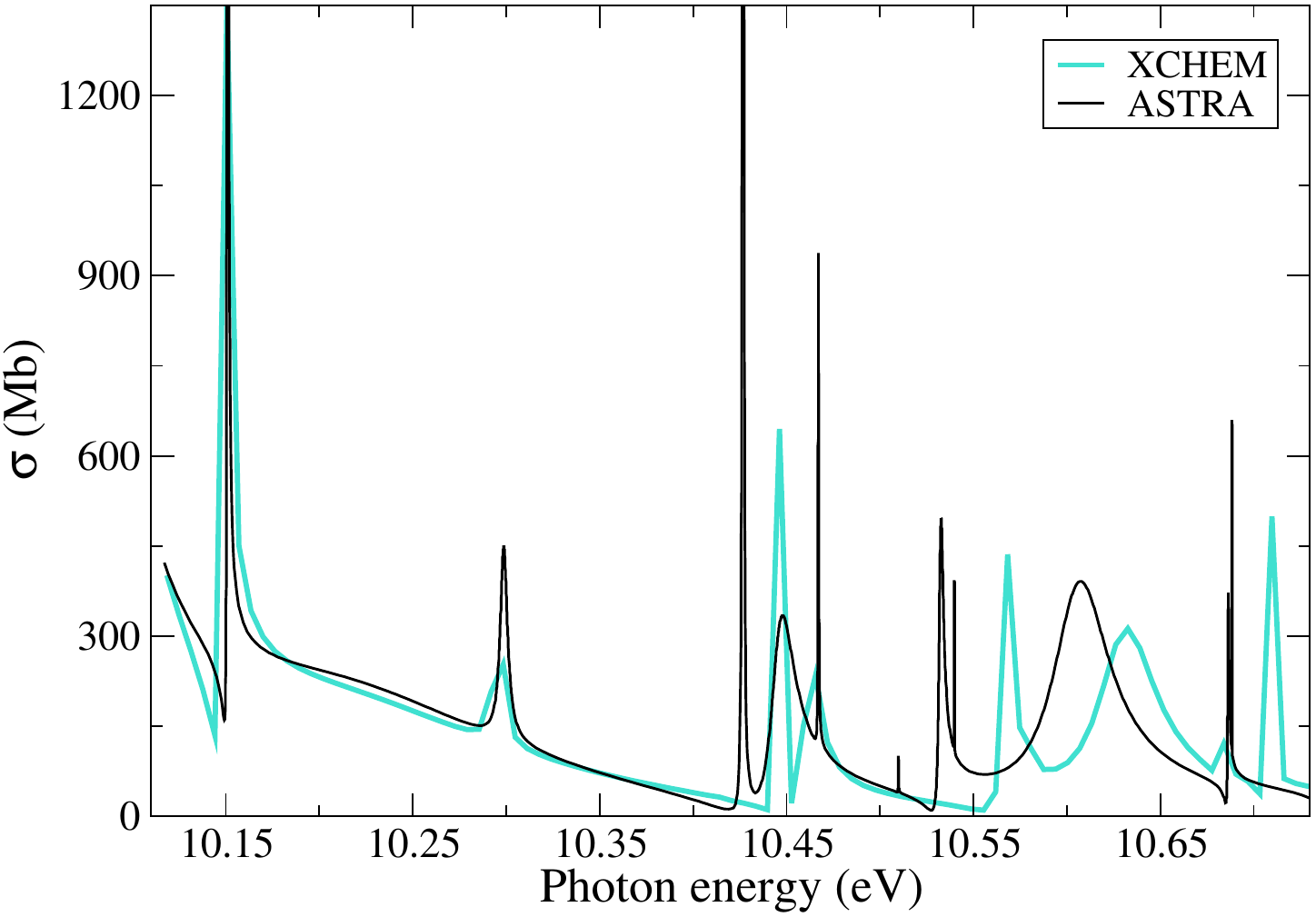}\hfill
\includegraphics[width=0.49\textwidth]{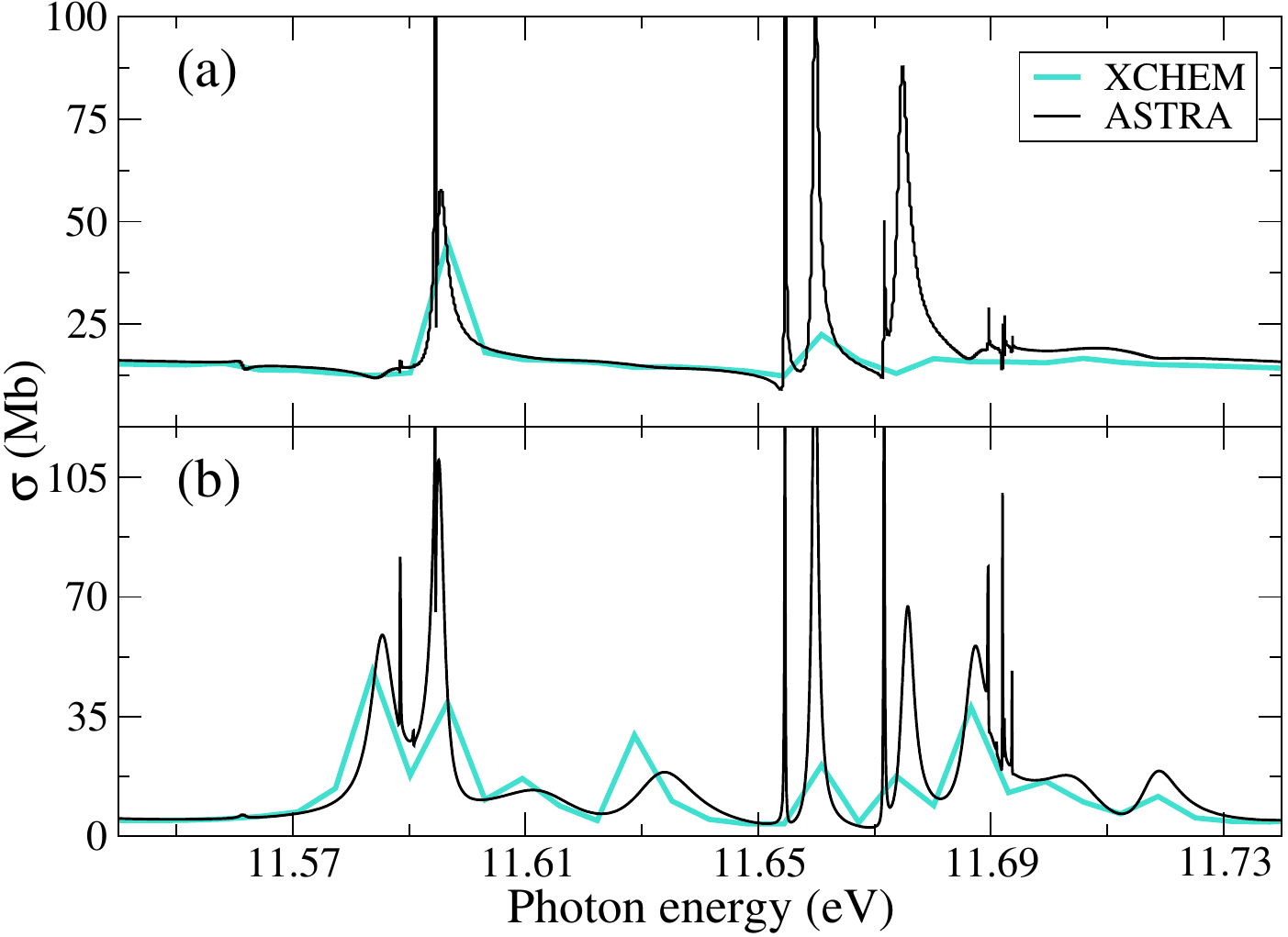}
\caption{\label{Fig:Pyrazine_TotCrossSec}\label{Fig:Pyrazine_PartialCS_Comparison}Pyrazine photoionization cross sections from \XCHEM{} (light thick curves) and \ASTRA{} (dark thin curves). Left: the $^2\mathrm B_{1g}$ partial cross section below the $^2\mathrm A_g$ threshold, where it is also the total cross section. Right: the $^2\mathrm B_{1g}$ [panel (a)] and $^2\mathrm A_g$ [panel (b)] partial cross sections between the $^2\mathrm A_g$ and $^2\mathrm B_{1u}$ thresholds.}
\end{widefigure}

The significance of this comparison is the consistency of independent molecular calculations at this scale. In \ASTRA{}, the same scattering construction used for the separable problem, the atomic test, and CO operates on a correlated polyatomic CC Hamiltonian assembled through ionic TDMs. The agreement with \XCHEM{} shows that this extension preserves the channel-resolved information needed for molecular photoionization.

\section{Conclusions and perspectives\label{sec:Conclusions}}

We have developed the PFG-VSM method for multichannel scattering states within the TDM close-coupling formulation of \ASTRA{}. The surface derivative is stationary with respect to errors in the interior wave function, and a pole-free auxiliary construction accommodates degenerate and closely spaced box eigenvalues. 

Through several comparisons, we establish complementary aspects of this method. Comparison with the exactly separable $\mathrm H_2^+$ problem tests the accuracy of PFG-VSM for a polycentric potential without many-electron approximations. The CO discretization study demonstrates the superior stability of variational matching: flux conservation remains accurate as radial coarsening degrades the direct match. Comparison with the results of an atomic code for Neon photoionization verifies the many-electron construction in the presence of degeneracies in the $D_{2h}$ representation. The excellent agreement between \ASTRA{} and \XCHEM{} for CO partial cross sections and MFPADs, evaluated on the same nominal close-coupling expansion, away from and across resonances, provides strong mutual validation of two independent molecular implementations. Their agreement for pyrazine extends this comparison to a substantially larger non-linear molecule.

PFG-VSM thus combines a stable scattering construction with the principal advantage of the TDM approach: the assembly of continuum couplings independently of the size of the ionic CI expansions. The resulting channel-resolved amplitudes bring partial cross sections and laboratory- and molecular-frame angular distributions within the same correlated molecular framework. In future works, this method will be applied to resolve in energy and angle the photoelectron distribution in attosecond pump-probe experiments, to compute angularly resolved photoemission time delays, and to determine elastic and inelastic electron-molecule scattering cross sections, which are a natural byproduct of the scattering-state calculations.

\begin{acknowledgments}
This work is supported by the DOE CAREER grant No. DE-SC0020311 and the DOE grant No. DE-SC0026377. We are grateful to Dr.~Zden\v{e}k Ma\v{s}\'in for fruitful discussions and suggestions on the use of the GBTOlib hybrid-integral library. We acknowledge generous allocation time at NERSC under the contract No. DE-AC02-05CH11231 and the award BES-ERCAP0024720, as well as time allocation from UCF Advanced Research Computing Center (ARCC). This material is based upon work supported by the National Science Foundation MPS-Ascend Postdoctoral Research Fellowship under Grant No. 2402225.
\end{acknowledgments}

\section*{Data availability}
The data that support the findings of this article are not publicly available. The data are available from the authors upon reasonable request.

\appendix

\section{Orientational average\label{app:OrientationalAverage}}

For the spin-free photoionization considered here, the incoming-wave scattering state with well-defined asymptotic parent-ion state $a$, total energy $E$, photoelectron direction $\hat\Omega$, and spin projection $\sigma$ is expanded in spin-adapted partial waves as~\cite{Newton}

\begin{equation}\label{eq:PlaneWavesInSphericalWaves}
\begin{split}
|\Psi_{a,E\hat{\Omega}\sigma}^-\rangle &=
\sum_{\ell m} i^{\ell -1}e^{-i\sigma_\ell(E)} Y_{\ell
m}^*(\hat{\Omega})\times\\
&\times\sum_{S\Sigma}C_{S_a\Sigma_a,\frac{1}{2}\sigma}^{S\Sigma}
|\Psi_{a,E\ell m;S\Sigma}^-\rangle.
\end{split}\end{equation}
These states are normalized as
\begin{equation}
\langle\Psi_{a,E\hat{\Omega}\sigma}^-|\Psi_{b,E'\hat{\Omega}'\sigma'}^-\rangle
= \delta_{ab} \delta_{\sigma\sigma'} \,k\,\delta(\vec{k}-\vec{k}'),
\end{equation}
where $\vec k=k_a\hat\Omega$ and $k_a=\sqrt{2(E-E_a)}$ is the asymptotic electron momentum. Equivalently, the normalization is $\delta_{ab}\delta_{\sigma\sigma'}\delta(E-E') \delta(\hat\Omega-\hat\Omega')$. This follows from closure of the spherical harmonics, $\delta(E-E')=k^{-1}\delta(k-k')$, and the Clebsch--Gordan orthogonality relation $\sum_{S\Sigma}C_{S_a\Sigma_a,\frac12\sigma}^{S\Sigma} C_{S_a\Sigma_a,\frac12\sigma'}^{S\Sigma}=\delta_{\sigma\sigma'}$. The interactions considered here conserve total spin and its projection, so we suppress $S$ and $\Sigma$ on the spin-adapted states from now on.

For a freely evolving wave packet $|\Psi(t)\rangle$, the molecular-frame distribution, unresolved in the spin projections of the ion and electron, is

\begin{equation}
\frac{dP_a}{dEd\Omega} = \sum_{\Sigma_a \sigma} \left|\langle
\Psi_{a,E\hat{\Omega}\sigma}^-|\Psi(t)\rangle\right|^2.
\end{equation}
The scattering states are eigenfunctions of the field-free, fixed-nuclei Hamiltonian, so this distribution is invariant during free evolution. For one-photon ionization from an unpolarized bound state $|g\rangle$ of spin $S_0$, averaging over the initial spin and summing over the final spins gives the differential cross section

\begin{equation}\label{eq:MFPADsGeneralFormula}
\frac{d\sigma_a^{\textsc{mf}}}{d\Omega} =
\frac{4\pi^2\omega}{(2S_0+1)c}\sum_{\Sigma_0\sigma\Sigma_a} \left|\langle
\Psi_{a,E\hat{\Omega}\sigma}^-|
\hat{\epsilon}\cdot\vec{d}|g\rangle\right|^2,
\end{equation}
where $\omega$ is the photon angular frequency, $c$ is the speed of light, $\hat{\epsilon}$ is the light polarization, and $\vec{d}=\sum_{i} q_i\vec{r}_i$ is the total electric dipole operator.

Substituting Eq.~\eqref{eq:PlaneWavesInSphericalWaves} into Eq.~\eqref{eq:MFPADsGeneralFormula} expresses the MFPAD in terms of dipole matrix elements from the ground state to scattering states with well-defined outgoing asymptotic spherical symmetry. The spin sums then contract. The spin-independent dipole operator conserves $S=S_0$, $\Sigma=\Sigma_0$, with matrix elements independent of $\Sigma_0$; the identity $\sum_{\alpha\beta\gamma}|C_{a\alpha,b\beta}^{c\gamma}|^2=2c+1$ ~\cite{Varshalovich1988} gives $\sum_{\Sigma_0\Sigma_a\sigma} |C_{S_a\Sigma_a,\frac12\sigma}^{S_0\Sigma_0}|^2=2S_0+1$, canceling the initial spin average. The MFPAD is thus a sesquilinear form in the partial-wave dipole matrix elements,

\begin{equation}\label{eq:MFPADBilinear}
\begin{split}
&\frac{d\sigma_a^{\textsc{mf}}}{d\Omega} = \frac{4\pi^2\omega}{c}
\sum_{\ell m\ell'm'} i^{\ell-\ell'}e^{-i(\sigma_\ell-\sigma_{\ell'})}\,
Y_{\ell m}^*(\hat{\Omega})Y_{\ell'm'}(\hat{\Omega})\times\\
&\times\langle g|d_\epsilon^\dagger|
\Psi_{a,E\ell'm'}^-\rangle^*\,
\langle g|d_\epsilon^\dagger|\Psi_{a,E\ell
m}^-\rangle.
\end{split}
\end{equation}
The spherical-harmonic product is reduced using~\cite{Varshalovich1988}

\begin{equation}\label{eq:YYYIntegral}
\langle
Y_{c\gamma}|Y_{b\beta}|Y_{a\alpha}\rangle
=\frac{\Pi_{ab}}{\sqrt{4\pi}\Pi_c}\,
C_{a0,b0}^{c0}\,C_{a\alpha,b\beta}^{c\gamma},
\end{equation}
where $\Pi_{ab\cdots}=\sqrt{(2a+1)(2b+1)\cdots}$ and $C_{a\alpha,b\beta}^{c\gamma}$ are Clebsch--Gordan coefficients. Indeed, the projection of $Y_{\ell m}^*Y_{\ell'm'}$ on $Y_{j\mu}$, $\int d\hat\Omega\,\big[Y_{\ell m}^{*}Y_{\ell'm'}\big]Y_{j\mu}^{*}$, is real and coincides with Eq.~\eqref{eq:YYYIntegral} for $(a,b,c)=(\ell,j,\ell')$, whence
\begin{equation}\label{eq:YstarYProduct}
Y_{\ell m}^{*}(\hat\Omega)\,Y_{\ell'm'}(\hat\Omega)=
\sum_{j\mu}\frac{\Pi_{j\ell}}{\sqrt{4\pi}\,\Pi_{\ell'}}\,
C_{\ell0,j0}^{\ell'0}\,C_{\ell m,j\mu}^{\ell'm'}\,Y_{j\mu}(\hat\Omega),
\end{equation}
with $\mu=m'-m$. The result is
\begin{equation}\label{eq:MFPADs}
\begin{split}
&\frac{d\sigma_a^{\textsc{mf}}}{d\Omega} = \frac{4\pi^2\omega}{c}
\sum_{j\mu} \frac{Y_{j\mu}(\hat{\Omega})}{\sqrt{4\pi}}\times\\
&\times\sum_{\ell\ell'}  e^{-i(\sigma_\ell-\sigma_{\ell'})}\,i^{\ell-\ell'}
C_{\ell 0, j0}^{\ell' 0}\frac{\Pi_{j\ell}}{\Pi_{\ell'}}\sum_{mm'}C_{\ell
m,j\mu}^{\ell' m'} \times\\
&\times\langle g|d_\epsilon^\dagger|
\Psi_{a,E\ell'm'}^-\rangle^*\,
\langle g|d_\epsilon^\dagger|\Psi_{a,E\ell
m}^-\rangle.
\end{split}
\end{equation}
Note that the parity-selection coefficient $C_{\ell0,j0}^{\ell'0}$ restricts the sum in Eq.~\eqref{eq:MFPADs} to $|\ell-\ell'|\leq j\leq\ell+\ell'$ with $\ell+\ell'+j$ even: a close-coupling expansion truncated at $\ell\leq\ell_{\mathrm{max}}$ therefore generates molecular-frame harmonics up to $j=2\ell_{\mathrm{max}}$.

To obtain the laboratory-frame distribution for randomly oriented molecules, we average the MFPAD over molecular orientations. Let the MF be defined by a rotation $\hat R$ from the LF, with Euler angles $\alpha$, $\beta$, and $\gamma$. The bound and scattering states of the rotated molecule are related to those of the unrotated molecule and ion by

\begin{equation}
|g\rangle = |g(\alpha,\beta,\gamma)\rangle \equiv
\hat{R}(\alpha,\beta,\gamma) |g_0\rangle,
\end{equation}
\begin{equation}\begin{split}
&|\Psi_{a,E\ell m\sigma}^-\rangle = |\Psi_{a(\alpha,\beta,\gamma),E\ell
m\sigma}^-\rangle =\\
&=\hat{R}(\alpha,\beta,\gamma)\sum_{\bar{m}}
|\Psi_{a_0,E\ell\bar{m}\sigma}^-\rangle D_{\bar{m}m}^{\ell}(\hat{R}^{-1}),
\end{split}\end{equation}
where $D^\ell_{mm'}(\hat R)=\langle\ell m|\hat R|\ell m'\rangle$ is a Wigner rotation matrix, and $g_0,a_0$ denote the neutral target and ion in a predefined, otherwise arbitrary, reference orientation in the LF. The sum over $\bar m$ is required because the photoelectron angular distribution always refers to the LF. The orientationally averaged cross section is

\begin{equation}
\frac{d\sigma_a^{\textsc{lf}}}{d\Omega}=\frac{1}{8\pi^2}\int d\mathcal{R}\,
\frac{d\sigma_a^{\textsc{mf}}(\alpha,\beta,\gamma)}{d\Omega},
\end{equation}
with $d\mathcal R=d\alpha\,\sin\beta\,d\beta\,d\gamma$, $\alpha,\gamma\in[0,2\pi]$, and $\beta\in[0,\pi]$ for integer angular momenta~\cite{Varshalovich1988}. Only the dipole-matrix-element product in Eq.~\eqref{eq:MFPADs} depends on molecular orientation.

For linear or circular polarization, choose the laboratory axis so that $d_\epsilon^\dagger$ is proportional to one spherical component $d_{1\xi}$, with $\xi=0$ or $\pm1$, respectively. The label $\xi$ refers to the reverse, ground-state-bra amplitude used in this derivation. Since $d_{1q}^\dagger=(-1)^q d_{1,-q}$, an absorption operator $d_\epsilon\propto d_{1q}$ has $\xi=-q$, up to a phase that cancels from the intensity. 
The transformation $\hat R^{-1}d_{1\xi}\hat R=\sum_\tau d_{1\tau} D^1_{\tau\xi}(\hat R^{-1})$~\cite{Varshalovich1988} expresses the amplitude in terms of reference-orientation matrix elements,
\begin{equation}\begin{split}
&\langle g|d_\epsilon^\dagger|\Psi_{a,E\ell m}^-\rangle=
\langle g_0|\hat{R}^{-1}(\alpha,\beta,\gamma)d_{1\xi}|\Psi_{a,E\ell
m}^-\rangle=\\
&=\sum_{\tau\bar{m}}
\langle g_0|d_{1\tau}|\Psi_{a_0,E\ell \bar{m}}^-\rangle
D_{\bar{m}m}^{\ell}(\hat{R}^{-1})\,D_{\tau \xi}^{1}(\hat{R}^{-1}).
\end{split}\end{equation}
We now use the unitarity of the Wigner matrices, $D_{mm'}^\ell(\hat R^{-1})=D_{m'm}^{\ell*}(\hat R)$, to express the result in terms of $\hat R$ rather than its inverse,
\begin{equation}\begin{split}
&\langle g|d_\epsilon^\dagger|\Psi_{a,E\ell m}^-\rangle=\\
&=\sum_{\tau\bar{m}}
\langle g_0|d_{1\tau}|\Psi_{a_0,E\ell \bar{m}}^-\rangle
D_{m\bar{m}}^{\ell*}(\hat{R})\,D_{\xi\tau }^{1*}(\hat{R}).
\end{split}\end{equation}
The product then is
\begin{equation}\begin{split}\label{eq:MEProduct}
&\langle g|d_\epsilon^\dagger|
\Psi_{a,E\ell'm'}^-\rangle^*\,
\langle g|d_\epsilon^\dagger|\Psi_{a,E\ell
m}^-\rangle=\\
&=\sum_{\tau\tau'\bar{m}\bar{m}'}
\langle g_0|d_{1\tau'}|\Psi_{a_0,E\ell' \bar{m}'}^-\rangle^*
\langle g_0|d_{1\tau}|\Psi_{a_0,E\ell \bar{m}}^-\rangle\times\\
&\times (-1)^{\tau-\xi}
D_{m\bar{m}}^{\ell*}(\hat{R})\,
D_{m'\bar{m}'}^{\ell'}(\hat{R})\,D_{-\xi,-\tau }^{1}(\hat{R})\,D_{\xi\tau'
}^{1}(\hat{R}),
\end{split}\end{equation}
where we used the relation $D_{mm'}^{\ell *}(\hat{R})=(-1)^{m'-m}D_{-m,-m'}^{\ell }(\hat{R})$ (see Eq.~4.4.2 of Ref.~\cite{Varshalovich1988}). The product between the two rank-one Wigner matrices is conveniently rewritten in terms of a single one using the Clebsch--Gordan series (see Eq.~4.6.1.1 of Ref.~\cite{Varshalovich1988})
\begin{equation}\begin{split}
&D_{mn}^{j}(\hat{R})\,D_{m'n'}^{j'}(\hat{R})=\\
&=\sum_{J=|j-j'|}^{j+j'}\sum_{MN}
C_{jm,j'm'}^{JM}C_{jn,j'n'}^{JN}D_{MN}^{J}(\hat{R}),
\end{split}\end{equation}
with which
\begin{equation}\begin{split}
&D_{-\xi,-\tau }^{1}(\hat{R})\,D_{\xi\tau' }^{1}(\hat{R})=\\
&=\sum_{J=0}^{2}\sum_{N}
C_{1-\xi,1\xi}^{J0}C_{1-\tau,1\tau'}^{JN}D_{0N}^{J}(\hat{R}).
\end{split}\end{equation}
The dynamical and geometrical factors independent of $\hat R$ can be taken outside the orientational integral. The remaining product $D_{m\bar m}^{\ell*}(\hat R)D_{m'\bar m'}^{\ell'}(\hat R)D_{0N}^{J}(\hat R)$ is averaged using Eq.~4.11.1.5 of Ref.~\cite{Varshalovich1988},

\begin{equation}\begin{split}
\frac{1}{8\pi^2}\int d\mathcal{R}\,
&D_{m\bar{m}}^{\ell*}(\hat{R})\,
D_{m'\bar{m}'}^{\ell'}(\hat{R})\,D_{0N}^{J}(\hat{R})=\\
&=\frac{C_{\ell' m',J 0}^{\ell m}C_{\ell' \bar{m}',J N}^{\ell
\bar{m}}}{2\ell+1}.
\end{split}\end{equation}
The orientational average of~\eqref{eq:MEProduct} then is
\begin{equation}\begin{split}\label{eq:AveragedMEProduct}
&\frac{1}{8\pi^2}\int
d\mathcal{R}\,\langle g|d_\epsilon^\dagger|
\Psi_{a,E\ell'm'}^-\rangle^*\,
\langle g|d_\epsilon^\dagger|\Psi_{a,E\ell
m}^-\rangle=\\
&=\sum_{\tau\tau'\bar{m}\bar{m}'}
\langle g_0|d_{1\tau'}|\Psi_{a_0,E\ell' \bar{m}'}^-\rangle^*
\langle g_0|d_{1\tau}|\Psi_{a_0,E\ell \bar{m}}^-\rangle\times\\
&\times
\frac{(-1)^{\tau-\xi+\ell'-\ell}}{\Pi_{\ell\ell'}}\,\sum_{J=0}^{2}C_{\ell m,J
0}^{\ell' m'}\times\\
&\times\sum_{N}C_{1-\xi,1\xi}^{J0}C_{1-\tau,1\tau'}^{JN}C_{\ell' \bar{m}',J
N}^{\ell \bar{m}},
\end{split}\end{equation}
where we used the symmetry relation of the Clebsch--Gordan coefficients $C_{\ell' m',J 0}^{\ell m}=(-1)^{\ell'-\ell}\, C_{\ell m,J 0}^{\ell' m'}\,\Pi_{\ell}/\Pi_{\ell'}$~\cite{Varshalovich1988}. The LFPAD now follows by replacing, in Eq.~\eqref{eq:MFPADs}, the product of the two dipole matrix elements by its orientational average~\eqref{eq:AveragedMEProduct}. The sum over the laboratory-frame projections $m$ and $m'$ then involves only the two Clebsch--Gordan coefficients that carry them, and it is evaluated with the orthogonality relation~\cite{Varshalovich1988}
\begin{equation}\label{eq:CGOrthogonality}
\sum_{mm'} C_{\ell m,J 0}^{\ell' m'}C_{\ell m,j\mu}^{\ell'
m'}=\frac{\Pi_{\ell'}^2}{\Pi_J^2}\delta_{jJ}\delta_{\mu 0},
\end{equation}
which follows from $3j$ orthogonality after applying the coefficient symmetries. The statistical weight is $2\ell'+1$, associated with the upper angular momentum; replacing it by $2\ell+1$ would change the $\ell\ne\ell'$ interference terms.

Equation~\eqref{eq:CGOrthogonality} restricts the distribution to $\mu=0$ and $j=J\leq2$, establishing axial symmetry about the linear-polarization direction ($\xi=0$) or the normal to the circular-polarization plane ($\xi=\pm1$). With $Y_{J0}(\hat\Omega)=\Pi_JP_J(\cos\theta)/\sqrt{4\pi}$, the statistical factors reduce to

\begin{equation}\label{eq:PiCollapse}
\frac{\Pi_{J\ell}}{\Pi_{\ell'}}\times
\frac{1}{\Pi_{\ell\ell'}}\times
\frac{\Pi_{\ell'}^2}{\Pi_J^2}\times
\frac{Y_{J0}(\hat\Omega)}{\sqrt{4\pi}}
=\frac{P_J(\cos\theta)}{4\pi},
\end{equation}
so these particular statistical factors cancel in the final distribution. Finally, the two rank-one coefficients of Eq.~\eqref{eq:AveragedMEProduct} are recast with the symmetry relations~\cite{Varshalovich1988}
\begin{equation}\label{eq:RankOneFlips}
\begin{aligned}
C_{1-\xi,1\xi}^{J0}
&=(-1)^{1+\xi-J}\,\frac{\Pi_J}{\Pi_1}\,C_{1\xi,J0}^{1\xi},\\
C_{1-\tau,1\tau'}^{JN}
&=(-1)^{1+\tau-J}\,\frac{\Pi_J}{\Pi_1}\,C_{1\tau,JN}^{1\tau'},
\end{aligned}
\end{equation}
whose phases combine with the factor $(-1)^{\tau-\xi}$ of Eq.~\eqref{eq:AveragedMEProduct} into $(-1)^{2+2\tau-2J}=1$, leaving the overall statistical factor $\Pi_J^2/\Pi_1^2=(2J+1)/3$. The factor $C_{1\xi,10}^{1\xi}$ makes the $J=1$ coefficient vanish for linear polarization. For circular polarization, molecular symmetry imposes an additional condition: the electric-dipole forward--backward asymmetry vanishes for an isotropic achiral sample and for a racemic mixture. A chiral sample with enantiomeric excess can have a nonzero coefficient, which reverses sign on helicity or enantiomer reversal~\cite{Ritchie1976, Goetz2016}. The partial LFPAD is

\begin{widetext}
\begin{equation}\label{eq:LFPADsFull}
\begin{split}
\frac{d\sigma_a^{\textsc{lf}}}{d\Omega} &= \sum_{J=0}^2
s_{Ja}\,P_{J}(\cos\theta),\qquad
s_{Ja}=\frac{\pi\omega}{3c}\,
(2J+1)\, C_{1\xi,J0}^{1\xi}
\sum_{\ell\ell'}  e^{-i(\sigma_\ell-\sigma_{\ell'})}\,i^{\ell-\ell'} C_{\ell
0, J0}^{\ell' 0} \,\times\\
&\times\sum_{\tau\tau'mm'N}
(-1)^{\ell'-\ell}\,C_{1\tau,JN}^{1\tau'}C_{\ell' m',J N}^{\ell
m}\,\langle g_0|d_{1\tau'}|\Psi_{a_0,E\ell' m'}^-\rangle^*
\langle g_0|d_{1\tau}|\Psi_{a_0,E\ell m}^-\rangle.
\end{split}
\end{equation}
\end{widetext}
The integral partial cross section takes contribution from $J=0$ only, $\sigma_a^{\textsc{lf}}=4\pi\,s_{0a}$, i.e.,
\begin{equation}\label{eq:IntegralPartialCS}
\begin{split}
\sigma_a^{\textsc{lf}} &= \frac{4\pi^2\omega}{3c}
\sum_{\ell m \tau}
\left|\langle g_0|d_{1\tau}|\Psi_{a_0,E\ell m}^-\rangle\right|^2,
\end{split}
\end{equation}
which is Eq.~\eqref{eq:SumPPCS} of the main text. In general, the LFPAD can be written as in Eq.~\eqref{eq:LFPADShape}, with the asymmetry parameters given by the ratios of the Legendre coefficients of Eq.~\eqref{eq:LFPADsFull},
\begin{equation}\label{eq:BetaParameters}
\beta_{Ja}=\frac{s_{Ja}}{s_{0a}},\qquad J=1,2.
\end{equation}

The polarization dependence of $s_{2a}$ is confined to $C_{1\xi,20}^{1\xi}$, and $C_{1\pm1,20}^{1\pm1}/C_{10,20}^{10}=-1/2$ recovers $\beta_{2a}^{\mathrm{circ}}=-\beta_{2a}^{\mathrm{lin}}/2$ for an isotropic ensemble in the electric-dipole approximation. A further check is ionization of an $s$ electron into a single $\ell=1$ channel: $\langle g_0|d_{1\tau}|\Psi^-_{a_0,E1\bar m}\rangle \propto C_{1\tau,1\bar m}^{00}$ by the Wigner--Eckart theorem, and Eq.~\eqref{eq:LFPADsFull} gives $s_{2a}=2s_{0a}$, or $\beta_{2a}^{\mathrm{lin}}=2$~\cite{CooperZare1968}. We have also verified Eqs.~\eqref{eq:MFPADs} and~\eqref{eq:LFPADsFull} against direct orientational quadrature for arbitrary complex amplitudes and $\xi=0,\pm1$.

\section{Analytical reference solution for \texorpdfstring{$\mathrm H_2^+$}{H2+}}\label{app:H2plusBenchmark}

We describe here the reference solution for $H_2^+$ scattering states used in Sec.~\ref{Subsec:H2plusResults}. Separation in prolate spheroidal coordinates reduces the fixed-nuclei two-center Coulomb problem to angular and radial equations~\cite{Bates1953,Power1973,Singor2023}. Their numerical solution yields the spheroidal eigenchannels and phases, from which we assemble the reaction matrix in the spherical-channel convention of the main text.

With the two protons on the $z$ axis at $z=\pm R/2$, and $r_{1,2}$ the distances of the electron from the two nuclei, the prolate spheroidal coordinates
\begin{equation}\label{eq:SpheroidalCoords}
\xi=\frac{r_1+r_2}{R}\in[1,\infty),
\qquad
\eta=\frac{r_1-r_2}{R}\in[-1,1],
\end{equation}
together with the azimuthal angle $\phi$, separate the fixed-nuclei Schr\"odinger equation at any energy~\cite{Bates1953,Power1973}. Writing the solution at photoelectron energy $\varepsilon=k^2/2$ as $\psi=X(\xi)\,S(\eta)\,e^{im\phi}$ and defining $\mu=|m|$, the Schr\"odinger equation splits into
\begin{align}
&\frac{d}{d\eta}\Big[(1-\eta^2)\frac{dS}{d\eta}\Big]
+\Big[A-c^2\eta^2-\frac{\mu^2}{1-\eta^2}\Big]S=0,
\label{eq:SpheroidalAngularODE}\\
&\frac{d}{d\xi}\Big[(\xi^2-1)\frac{dX}{d\xi}\Big]
+\Big[c^2\xi^2+\alpha\,\xi-A-\frac{\mu^2}{\xi^2-1}\Big]X=0,
\label{eq:SpheroidalRadialODE}
\end{align}
coupled only through the separation constant $A$, with the two dimensionless parameters
$c=kR/2$ and $\alpha=2R$.
Equation~\eqref{eq:SpheroidalAngularODE} is the standard spheroidal angular equation~\cite{Flammer1957}. Expanding $S(\eta)=\sum_\ell a_\ell\,\bar P_\ell^\mu(\eta)$ over normalized associated Legendre functions turns it into a real symmetric eigenproblem,
\begin{equation}\label{eq:SpheroidalAngularEig}
\sum_{\ell'}\Big[\ell(\ell+1)\,\delta_{\ell\ell'}
+c^2\,\langle\ell|\eta^2|\ell'\rangle\Big]\,a_{\ell'}=A\,a_\ell,
\end{equation}
with $\langle\ell|\eta^2|\ell\rangle =\big(a^{(\mu)}_{\ell}\big)^2+\big(a^{(\mu)}_{\ell-1}\big)^2$ and $\langle\ell|\eta^2|\ell\pm2\rangle=a^{(\mu)}_{\ell_<}a^{(\mu)}_{\ell_<+1}$ ($\ell_<$ the smaller of $\ell$, $\ell'$), where $a^{(\mu)}_\ell=\sqrt{(\ell+1-\mu)(\ell+1+\mu)/[(2\ell+1)(2\ell+3)]}$ are the coefficients of the Legendre three-term recursion $\eta\,\bar P^\mu_\ell=a^{(\mu)}_\ell\bar P^\mu_{\ell+1}+a^{(\mu)}_{\ell-1}\bar P^\mu_{\ell-1}$. The eigenvalues give $A_{\mu q}(c)$, and the orthonormal eigenvectors $c_{\ell q}$ expand the spheroidal channels $q=\mu,\mu+1,\ldots$ over the spherical partial waves of Eq.~\eqref{eq:PWCDefinition}, with $q\to\ell$ as $c\to0$.
Reflection in the plane perpendicular to the molecular axis has parity $(-1)^{q-\mu}$, and inversion has parity $g/u=(-1)^q$.
At $c\lesssim1.8$, the spheroidal Legendre coefficients decay superexponentially in $|\ell-q|$, so a modest angular margin beyond the largest tested $\ell$ suffices.
For the radial problem, the substitution $u(\xi)=\sqrt{\xi^2-1}\,X(\xi)$ gives the exact normal-form coefficient $Q(\xi)=(c^2\xi^2+\alpha\xi-A)/(\xi^2-1) +(1-\mu^2)/(\xi^2-1)^2$. Its large-$\xi$ expansion is
\begin{equation}\label{eq:SpheroidalNormalForm}
u''+\Big[c^2+\frac{\alpha}{\xi}+\frac{c^2-A}{\xi^{2}}
+\frac{\alpha}{\xi^{3}}+O\big(\xi^{-4}\big)\Big]u=0.
\end{equation}
Comparison with the radial Coulomb equation in the variable $\rho=c\,\xi=k\frac{R}{2}\xi\to kr$ identifies the asymptotic reference: a Coulomb wave with charge 2 and the generally \emph{non-integer} angular momentum $L_q$,
\begin{equation}\label{eq:SpheroidalL}
L_q(L_q+1)=A_{\mu q}(c)-c^2,
\quad
L_q=-\tfrac12+\sqrt{\tfrac14+A_{\mu q}-c^2}\,,
\end{equation}
which reduces to $q$ (the corresponding spherical $\ell$) as $c\to0$. For matching and phase conversion, we use the real order $L_q^{\mathrm{ref}}=L_q$ when $A_{\mu q}-c^2\geq-1/4$, and $L_q^{\mathrm{ref}}=-1/2$ otherwise. The radial equation is unchanged.

The regular solution, which behaves as $X\sim(\xi^2-1)^{\mu/2}$ at the singular endpoint, is integrated outward from $\xi=1$ and decomposed at a large radius $\xi_M$ over the Coulomb pair of order $L_q^{\mathrm{ref}}$, 
\begin{align}
u&=P\,F_{L_q^{\mathrm{ref}}}+Q\,G_{L_q^{\mathrm{ref}}},
\quad\delta_q(\xi_M)=\operatorname{atan2}(Q,P),
\label{eq:SpheroidalPhaseMatch}\\
\frac{du}{d\rho}
&=\frac{1}{c}\frac{du}{d\xi},\qquad
\frac{du}{d\xi}
=\frac{\xi X+(\xi^2-1)X'}{\sqrt{\xi^2-1}}.
\nonumber
\end{align}
The asymptotic phase $\delta_q$ is obtained by fitting $\delta_q(\xi_M)=\delta_q+b_1/\xi_M+b_2/\xi_M^2$ at three matching radii and extrapolating to $\xi_M\to\infty$; the $1/\xi_M$ term accounts for the centrifugal mismatch when $L_q^{\mathrm{ref}}\ne L_q$.
Since $\rho=c\xi\to kr$ and $r-\frac R2\xi\to0$, the channel solution has the asymptotic form, up to normalization,
\begin{equation}\label{eq:SpheroidalAsym}
\psi_{q}\;\underset{r\to\infty}{\longrightarrow}\;
\Big[\sum_{\ell}c_{\ell q}\,X_{\ell m}(\hat r)\Big]\,
\frac{\sin\!\big[\theta_{L_q^{\mathrm{ref}}}(r)+\delta_q\big]}{kr},
\end{equation}
Here $\theta_L$ is the Coulomb phase of Eq.~\eqref{eq:CoulombPhase} for $Z=Z_1+Z_2$, continued to noninteger order through $\sigma_L=\arg\Gamma(L+1-iZ/k)$. We express the common radial wave relative to each spherical channel's phase $\theta_\ell$ as
\begin{equation}\label{eq:SpheroidalTrig}
\sin\big(\theta_{L_q^{\mathrm{ref}}}+\delta_q\big)
=\cos\beta_{\ell q}\,\sin\theta_\ell
+\sin\beta_{\ell q}\,\cos\theta_\ell,
\end{equation}
with $\beta_{\ell q}=\delta_q+(\ell-L_q^{\mathrm{ref}})\pi/2+\sigma_{L_q^{\mathrm{ref}}}-\sigma_{\ell}$,
where $\sin\theta_\ell$ and $\cos\theta_\ell$ are proportional to the regular and irregular Coulomb functions $\mathcal F_\ell$ and $\mathcal G_\ell$. Their energy-normalization factor, Eq.~\eqref{eq:ExponentialCombinations}, is common to all channels of this single-threshold problem and drops out of $\mathbf K$. The coefficient pair of Eq.~\eqref{eq:AsymptoticExpansion} is thus
\begin{equation}\label{eq:SpheroidalABex}
\mathbf A^{\mathrm{ex}}_{\ell q}=c_{\ell q}\cos\beta_{\ell q},
\qquad
\mathbf B^{\mathrm{ex}}_{\ell q}=c_{\ell q}\sin\beta_{\ell q}.
\end{equation}
From $(\mathbf A^{\mathrm{ex}},\mathbf B^{\mathrm{ex}})$ we obtain the reaction matrix and eigenphases as in Eq.~\eqref{eq:KMatrix}, within each $\mu$ block, assembling the blocks diagonally within each irreducible representation. The phases are compared with those from \ASTRA{} modulo $\pi$, after adapting its phases to Eq.~\eqref{eq:KMatrix}.

\bibliography{ASTRA_2026_Oct16}

@article{Joseph2023,
author = {Joseph, Jennifer and Holzmeier, Fabian and Bresteau, David and Ruchon, Thierry and Houver, Jean-Christophe and Lucchese, Robert R. and Dowek, Danielle},
doi = {10.1140/epjs/s11734-023-00815-7},
issn = {1951-6355},
journal = {Eur. Phys. J. Spec. Top.},
month = {apr},
publisher = {Springer Berlin Heidelberg},
title = {{Angularly resolved photoionization dynamics in atoms and molecules combining temporally and spectrally resolved experiments at ATTOLab and Synchrotron SOLEIL}},
url = {https://doi.org/10.1140/epjs/s11734-023-00815-7},
volume = {232},
year = {2023},
pages = {2031--2044}
}

@article{Dowek2022,
author = {Dowek, Danielle and Decleva, Piero},
doi = {10.1039/D2CP02725A},
issn = {1463-9076},
journal = {Phys. Chem. Chem. Phys.},
number = {40},
pages = {24614--24654},
pmid = {36205774},
publisher = {Royal Society of Chemistry},
title = {{Trends in angle-resolved molecular photoelectron spectroscopy}},
url = {http://xlink.rsc.org/?DOI=D2CP02725A},
volume = {24},
year = {2022}
}

@article{Rist2021,


author = {Rist, Jonas and others},
doi = {10.1038/s41467-021-26994-2},
issn = {2041-1723},
journal = {Nat. Commun.},
month = {nov},
number = {1},
pages = {6657},
pmid = {34789736},
title = {{Measuring the photoelectron emission delay in the molecular frame}},
url = {https://www.nature.com/articles/s41467-021-26994-2},
volume = {12},
year = {2021}
}

@article{Borras2021,
author = {Borr{\`{a}}s, Vicent J. and Gonz{\'{a}}lez-V{\'{a}}zquez, Jes{\'{u}}s and Argenti, Luca and Mart{\'{i}}n, Fernando},
doi = {10.1021/acs.jctc.1c00480},
issn = {1549-9618},
journal = {J. Chem. Theory Comput.},
month = {oct},
number = {10},
pages = {6330--6339},
pmid = {34528784},
title = {Molecular-frame photoelectron angular distributions of {CO} in the vicinity of {Feshbach} resonances: An {XCHEM} approach},
url = {https://pubs.acs.org/doi/10.1021/acs.jctc.1c00480 http://www.ncbi.nlm.nih.gov/pubmed/34528784},
volume = {17},
year = {2021}
}

@article{Koch2019,
author = {Koch, Christiane P and Lemeshko, Mikhail and Sugny, Dominique},
doi = {10.1103/RevModPhys.91.035005},
issn = {0034-6861},
journal = {Rev. Mod. Phys.},
month = {sep},
number = {3},
pages = {035005},
publisher = {American Physical Society},
title = {{Quantum control of molecular rotation}},
url = {https://doi.org/10.1103/RevModPhys.91.035005 https://link.aps.org/doi/10.1103/RevModPhys.91.035005},
volume = {91},
year = {2019}
}

@article{Fukuzawa2019,
author = {Fukuzawa, Hironobu and Yamada, Syuhei and Sakakibara, Yuta and Tachibana, Tetsuya and Ito, Yuta and Takanashi, Tsukasa and Nishiyama, Toshiyuki and Sakai, Tsukasa and Nagaya, Kiyonobu and Saito, Norio and Oura, Masaki and Stener, Mauro and Decleva, Piero and Ueda, Kiyoshi},
doi = {10.1063/1.5115801},
issn = {0021-9606},
journal = {J. Chem. Phys.},
month = {sep},
number = {10},
pmid = {31521098},
publisher = {AIP Publishing, LLC},
title = {{Probing gaseous molecular structure by molecular-frame photoelectron angular distributions}},
url = {https://doi.org/10.1063/1.5115801 https://pubs.aip.org/jcp/article/151/10/104302/197665/Probing-gaseous-molecular-structure-by-molecular},
volume = {151},
year = {2019},
pages = {104302}
}

@article{Ueda2019,
author = {Ueda, Kiyoshi and others},
doi = {10.1088/1361-6455/ab26d7},
issn = {0953-4075},
journal = {J. Phys. B: At. Mol. Opt. Phys.},
month = {sep},
number = {17},
pages = {171001},
title = {{Roadmap on photonic, electronic and atomic collision physics: I. Light–matter interaction}},
url = {https://iopscience.iop.org/article/10.1088/1361-6455/ab26d7},
volume = {52},
year = {2019}
}

@article{Goetz2016,


author = {Goetz, R. E. and Isaev, T. A. and Nikoobakht, B. and Berger, R. and Koch, C. P.},
doi = {10.1063/1.4973456},
issn = {00219606},
journal = {J. Chem. Phys.},
month = {jun},
number = {2},
pmid = {28088152},
title = {{Theoretical description of circular dichroism in photoelectron angular distributions of randomly oriented chiral molecules after multi-photon photoionization}},
url = {http://dx.doi.org/10.1063/1.4973456 http://arxiv.org/abs/1606.04436},
volume = {146},
year = {2017},
pages = {024306}
}

@article{Reid2012,
author = {Reid, Katharine L.},
doi = {10.1080/00268976.2011.640292},
issn = {0026-8976},
journal = {Mol. Phys.},
month = {feb},
number = {3},
pages = {131--147},
title = {{Photoelectron angular distributions: developments in applications to isolated molecular systems}},
url = {http://www.tandfonline.com/doi/abs/10.1080/00268976.2011.640292},
volume = {110},
year = {2012}
}

@article{Holmegaard2010,
author = {Holmegaard, Lotte and Hansen, Jonas L. and Kalh{\o}j, Line and {Louise Kragh}, Sofie and Stapelfeldt, Henrik and Filsinger, Frank and K{\"{u}}pper, Jochen and Meijer, Gerard and Dimitrovski, Darko and Abu-samha, Mahmoud and Martiny, Christian P. J. and {Bojer Madsen}, Lars},
doi = {10.1038/nphys1666},
issn = {1745-2473},
journal = {Nat. Phys.},
month = {jun},
number = {6},
pages = {428--432},
publisher = {Nature Publishing Group},
title = {{Photoelectron angular distributions from strong-field ionization of oriented molecules}},
url = {https://www.nature.com/articles/nphys1666},
volume = {6},
year = {2010}
}

@incollection{Seideman2005,
author = {Seideman, Tamar and Hamilton, Edward},
booktitle = {Advances in Atomic, Molecular and Optical Physics},
publisher = {Academic Press},
doi = {10.1016/S1049-250X(05)52006-8},
isbn = {9780120038527},
issn = {1049250X},
pages = {289--329},
title = {Nonadiabatic alignment by intense pulses. {Concepts}, theory, and directions},
url = {https://linkinghub.elsevier.com/retrieve/pii/S1049250X05520068},
volume = {52},
year = {2005}
}

@article{Dowek2004,
author = {Dowek, D. and Lebech, M. and Houver, J.C. and Lucchese, R.R.},
doi = {10.1016/j.elspec.2004.06.012},
issn = {03682048},
journal = {J. Electron Spectrosc. Relat. Phenom.},
month = {dec},
number = {2-3},
pages = {211--227},
title = {{Photoemission in the molecular frame using the vector correlation approach: from valence to inner-valence shell ionization}},
url = {https://linkinghub.elsevier.com/retrieve/pii/S0368204804003391},
volume = {141},
year = {2004}
}

@article{Lafosse2002,
author = {Lafosse, A. and Brenot, J. C. and Guyon, P. M. and Houver, J. C. and Golovin, A. V. and Lebech, M. and Dowek, D. and Lin, P. and Lucchese, R. R.},
doi = {10.1063/1.1512650},
issn = {0021-9606},
journal = {J. Chem. Phys.},
month = {nov},
number = {18},
pages = {8368--8384},
title = {{Vector} correlations in dissociative photoionization of {$\mathrm{O}_2$} in the 20--28 {eV} range. {II}. {Polar} and azimuthal dependence of the molecular frame photoelectron angular distribution},
url = {https://pubs.aip.org/jcp/article/117/18/8368/438796/Vector-correlations-in-dissociative},
volume = {117},
year = {2002}
}

@article{Lafosse2000,
author = {Lafosse, A. and Lebech, M. and Brenot, J. C. and Guyon, P. M. and Jagutzki, O. and Spielberger, L. and Vervloet, M. and Houver, J. C. and Dowek, D.},
doi = {10.1103/PhysRevLett.84.5987},
issn = {0031-9007},
journal = {Phys. Rev. Lett.},
month = {jun},
number = {26},
pages = {5987--5990},
pmid = {10991105},
title = {Vector correlations in dissociative photoionization of diatomic molecules in the {VUV} range: Strong anisotropies in electron emission from spatially oriented {NO} molecules},
url = {https://link.aps.org/doi/10.1103/PhysRevLett.84.5987},
volume = {84},
year = {2000}
}

@article{Larsen2000,
author = {Larsen, Jakob Juul and Hald, Kasper and Bjerre, Nis and Stapelfeldt, Henrik and Seideman, Tamar},
doi = {10.1103/PhysRevLett.85.2470},
issn = {0031-9007},
journal = {Phys. Rev. Lett.},
month = {sep},
number = {12},
pages = {2470--2473},
pmid = {10978084},
title = {Three dimensional alignment of molecules using elliptically polarized laser fields},
url = {https://link.aps.org/doi/10.1103/PhysRevLett.85.2470},
volume = {85},
year = {2000}
}

@article{Golovin1997,
author = {Golovin, A. V. and Heiser, F. and Quayle, C. J. K. and Morin, P. and Simon, M. and Gessner, O. and Guyon, P.-M. and Becker, U.},
doi = {10.1103/PhysRevLett.79.4554},
issn = {0031-9007},
journal = {Phys. Rev. Lett.},
month = {dec},
number = {23},
pages = {4554--4557},
title = {Observation of site-specific electron emission in the decay of superexcited {$\mathrm{O}_2$}},
url = {https://link.aps.org/doi/10.1103/PhysRevLett.79.4554},
volume = {79},
year = {1997}
}

@article{Randazzo2023,
author = {Randazzo, Juan Mart{\'{i}}n and Marante, Carlos and Chattopadhyay, Siddhartha and Schneider, Barry I. and Olsen, Jeppe and Argenti, Luca},
doi = {10.1103/PhysRevResearch.5.043115},
issn = {2643-1564},
journal = {Phys. Rev. Research},
month = {nov},
number = {4},
pages = {043115},
publisher = {American Physical Society},
title = {{ASTRA: Transition-density-matrix approach to molecular ionization}},
url = {https://link.aps.org/doi/10.1103/PhysRevResearch.5.043115},
volume = {5},
year = {2023}
}

@book{Newton,
author = {Newton, R G},
isbn = {9783642881282},
publisher = {Springer Berlin Heidelberg},
series = {Theoretical and Mathematical Physics},
title = {{Scattering Theory of Waves and Particles}},
url = {https://books.google.com/books?id=KWTyCAAAQBAJ},
year = {2013}
}

@article{McCurdy1979,
   author = {C. W. McCurdy and T. N. Rescigno},
   doi = {10.1103/PhysRevA.20.2346},
   issn = {0556-2791},
   issue = {6},
   journal = {Phys. Rev. A},
   month = {12},
   pages = {2346-2351},
   title = {Basis-set calculation of {Siegert} eigenvalues: Partial resonance widths},
   volume = {20},
   url = {https://link.aps.org/doi/10.1103/PhysRevA.20.2346},
   year = {1979},
}

@article{Siegert1939,
   author = {A. J. F. Siegert},
   doi = {10.1103/PhysRev.56.750},
   issn = {0031-899X},
   issue = {8},
   journal = {Phys. Rev.},
   month = {10},
   pages = {750-752},
   title = {On the Derivation of the Dispersion Formula for Nuclear Reactions},
   volume = {56},
   url = {https://link.aps.org/doi/10.1103/PhysRev.56.750},
   year = {1939},
}

@article{Lucchese1986,
author = {Lucchese, Robert R. and Takatsuka, Kazuo and McKoy, Vincent},
title = {Applications of the {Schwinger} variational principle to electron-molecule collisions and molecular photoionization},
journal = {Phys. Rep.},
volume = {131},
pages = {147--221},
year = {1986}
}

@article{Gianturco1994,
author = {Gianturco, F. A. and Lucchese, R. R. and Sanna, N.},
title = {Calculation of low-energy elastic cross sections for electron-{CF$_4$} scattering},
journal = {J. Chem. Phys.},
volume = {100},
pages = {6464--6471},
year = {1994}
}

@article{Natalense1999,
author = {Natalense, Alexandra P. P. and Lucchese, Robert R.},
title = {Cross section and asymmetry parameter calculation for sulfur 1s photoionization of {SF$_6$}},
journal = {J. Chem. Phys.},
volume = {111},
pages = {5344--5348},
year = {1999}
}

@article{McCurdy1987,
author = {McCurdy, C. W. and Rescigno, T. N. and Schneider, B. I.},
title = {Interrelation between variational principles for scattering amplitudes and generalized {$R$}-matrix theory},
journal = {Phys. Rev. A},
volume = {36},
pages = {2061--2066},
year = {1987},
doi = {10.1103/PhysRevA.36.2061}
}

@incollection{Rescigno1995,
author = {Rescigno, T. N. and McCurdy, C. W. and Orel, A. E. and Lengsfield III, B. H.},
title = {The complex {K}ohn variational method},
booktitle = {Computational Methods for Electron--Molecule Collisions},
editor = {Huo, W. M. and Gianturco, F. A.},
publisher = {Plenum Press},
address = {New York},
pages = {1--44},
year = {1995}
}

@article{Tennyson2010,
author = {Tennyson, Jonathan},
title = {Electron--molecule collision calculations using the {$R$}-matrix method},
journal = {Phys. Rep.},
volume = {491},
pages = {29--76},
year = {2010}
}

@article{Masin2020,
author = {Ma\v{s}\'in, Zden\v{e}k and Benda, Jakub and Gorfinkiel, Jimena D. and Harvey, Alex G. and Tennyson, Jonathan},
title = {{UKRmol+}: A suite for modelling electronic processes in molecules interacting with electrons, positrons and photons using the {$R$}-matrix method},
journal = {Comput. Phys. Commun.},
volume = {249},
pages = {107092},
year = {2020}
}

@misc{GBTOlib-4,
author = {Ma\v{s}\'in, Zden\v{e}k and Benda, Jakub and Gorfinkiel, Jimena D.},title = {{GBTOlib}: a high-performance library for evaluation of molecular integrals},howpublished = {Zenodo},note = {Version 4.0.0},doi = {10.5281/zenodo.18328470},year = {2026}}

@book{Burke2011,
author = {Burke, P. G.},
title = {{$R$}-Matrix Theory of Atomic Collisions: Application to Atomic, Molecular and Optical Processes},
publisher = {Springer},
address = {Berlin},
year = {2011}
}

@article{Marante2017,
author = {Marante, Carlos and Klinker, Markus and Corral, In\'es and Gonz\'alez-V\'azquez, Jes\'us and Argenti, Luca and Mart\'in, Fernando},
title = {Hybrid-basis close-coupling interface to quantum chemistry packages for the treatment of ionization problems},
journal = {J. Chem. Theory Comput.},
volume = {13},
pages = {499--514},
year = {2017},
doi = {10.1021/acs.jctc.6b00907}
}

@article{Bachau2001,
author = {Bachau, H. and Cormier, E. and Decleva, P. and Hansen, J. E. and Mart\'in, F.},
title = {Applications of {B}-splines in atomic and molecular physics},
journal = {Rep. Prog. Phys.},
volume = {64},
pages = {1815--1943},
year = {2001}
}

@article{Thompson1985,
author = {Thompson, I. J. and Barnett, A. R.},
title = {{COULCC}: A continued-fraction algorithm for {C}oulomb functions of complex order with complex arguments},
journal = {Comput. Phys. Commun.},
volume = {36},
pages = {363--372},
year = {1985},
doi = {10.1016/0010-4655(85)90025-6}
}

@article{Seaton1983,
author = {Seaton, M. J.},
title = {Quantum defect theory},
journal = {Rep. Prog. Phys.},
volume = {46},
pages = {167--257},
year = {1983}
}

@article{Bates1953,
author = {Bates, D. R. and Ledsham, K. and Stewart, A. L.},
title = {Wave functions of the hydrogen molecular ion},
journal = {Philos. Trans. R. Soc. London, Ser. A},
volume = {246},
pages = {215--240},
year = {1953}
}

@article{Power1973,
author = {Power, J. D.},
title = {Fixed nuclei two-centre problem in quantum mechanics},
journal = {Philos. Trans. R. Soc. London, Ser. A},
volume = {274},
pages = {663--697},
year = {1973}
}

@book{Flammer1957,
author = {Flammer, C.},
title = {Spheroidal Wave Functions},
publisher = {Stanford University Press},
address = {Stanford, CA},
year = {1957}
}

@book{Varshalovich1988,
author = {Varshalovich, D. A. and Moskalev, A. N. and Khersonskii, V. K.},
title = {Quantum Theory of Angular Momentum},
publisher = {World Scientific},
address = {Singapore},
year = {1988}
}

@article{Dunning1989,
author = {Dunning, Thom H.},
title = {Gaussian basis sets for use in correlated molecular calculations. {I}. {The} atoms boron through neon and hydrogen},
journal = {J. Chem. Phys.},
volume = {90},
pages = {1007--1023},
year = {1989}
}

@article{Kendall1992,
author = {Kendall, Rick A. and Dunning, Thom H. and Harrison, Robert J.},
title = {Electron affinities of the first-row atoms revisited. {Systematic} basis sets and wave functions},
journal = {J. Chem. Phys.},
volume = {96},
pages = {6796--6806},
year = {1992}
}

@article{Dalton2014,
author = {Aidas, Kestutis and others},
title = {The {D}alton quantum chemistry program system},
journal = {WIREs Comput. Mol. Sci.},
volume = {4},
pages = {269--284},
year = {2014}
}

@article{Olsen1988,
author = {Olsen, Jeppe and Roos, Bj{\"o}rn O. and J{\o}rgensen, Poul and Jensen, Hans J{\o}rgen Aa.},
title = {Determinant based configuration interaction algorithms for complete and restricted configuration interaction spaces},
journal = {J. Chem. Phys.},
volume = {89},
pages = {2185--2192},
year = {1988}
}

@article{Olsen2000,
author = {Olsen, Jeppe},
title = {The initial implementation and applications of a general active space coupled cluster method},
journal = {J. Chem. Phys.},
volume = {113},
pages = {7140--7148},
year = {2000}
}

@article{CooperZare1968,
author = {Cooper, John and Zare, Richard N.},
title = {Angular distribution of photoelectrons},
journal = {J. Chem. Phys.},
volume = {48},
pages = {942--943},
year = {1968}
}

@article{Ritchie1976,
 author = {Ritchie, Burke},
 title = {Theory of the angular distribution of photoelectrons ejected from optically active molecules and molecular negative ions},
 journal = {Phys. Rev. A}, volume = {13}, pages = {1411--1415},
 year = {1976}, doi = {10.1103/PhysRevA.13.1411}
}

@article{Borras2024,
author = {Borr{\`a}s, Vicent J. and Fern{\'a}ndez-Mil{\'a}n, Pedro and Argenti, Luca and Gonz{\'a}lez-V{\'a}zquez, Jes{\'u}s and Mart{\'i}n, Fernando},
title = {Photoionization cross sections and photoelectron angular distributions of molecules with {XCHEM-2.0}},
journal = {Comput. Phys. Commun.},
volume = {296},
pages = {109033},
year = {2024},
doi = {10.1016/j.cpc.2023.109033},
url = {https://doi.org/10.1016/j.cpc.2023.109033}
}

@article{Toffoli2024,
  author = {Toffoli, Daniele and Coriani, Sonia and Stener, Mauro and Decleva, Piero},
  title = {{Tiresia}: A code for molecular electronic continuum states and photoionization},
  journal = {Comput. Phys. Commun.},
  volume = {297},
  pages = {109038},
  year = {2024},
  doi = {10.1016/j.cpc.2023.109038}
}

@article{Chundayil2024,
  author = {Chundayil, Hareesh and Majety, Vinay P. and Scrinzi, Armin},
  title = {The hybrid anti-symmetrized coupled channels method ({haCC}) for the {tRecX} code},
  journal = {Comput. Phys. Commun.},
  volume = {303},
  pages = {109279},
  year = {2024},
  doi = {10.1016/j.cpc.2024.109279}
}

@article{Martin2026,
  author = {Mart{\'i}n, Fernando and Benda, Jakub and Gorfinkiel, Jimena D. and Ma{\v{s}}{\'i}n, Zden{\v{e}}k and Rego, Laura and Scrinzi, Armin},
  title = {Tutorial: theoretical methods for attosecond molecular ionization and dynamics},
  journal = {Commun. Phys.},
  volume = {9},
  pages = {182},
  year = {2026},
  doi = {10.1038/s42005-026-02671-y}
}

@article{Singor2023,
  author = {Singor, Adam and Savage, Jeremy S. and Bray, Igor and Schneider, Barry I. and Fursa, Dmitry V.},
  title = {Continuum solutions to the two-center {Coulomb} problem in prolate spheroidal coordinates},
  journal = {Comput. Phys. Commun.},
  volume = {282},
  pages = {108514},
  year = {2023},
  doi = {10.1016/j.cpc.2022.108514}
}

@article{Luo2024,
  title = {Influence of final state interactions in attosecond photoelectron interferometry},
  author = {Luo, S. and Weissenbilder, R. and Laurell, H. and Bello, R. Y. and Marante, C. and Ammitzb{\"o}ll, M. and Neori{\v{c}}i{\'{c}}, L. and Ljungdahl, A. and Squibb, R. J. and Feifel, R. and Gisselbrecht, M. and Arnold, C. L. and Mart{\'i}n, F. and Lindroth, E. and Argenti, L. and Busto, D. and L'Huillier, A.},
  journal = {Phys. Rev. Research},
  volume = {6},
  pages = {043271},
  year = {2024},
  doi = {10.1103/PhysRevResearch.6.043271}
}

@article{Wood2026,
  title = {Attosecond photoelectron interferometry of resonant structures between the {Ar} spin-orbit-split thresholds},
  author = {Wood, J. K. and Marante, C. and Biswas, D. and Argenti, L. and Sandhu, A.},
  journal = {Phys. Rev. Research},
  volume = {8},
  pages = {033290},
  year = {2026},
  doi = {10.1103/hz3c-jn2g}
}

\end{document}